\documentclass[%
 reprint,
 amsmath,amssymb,
 aps,
]{revtex4-2}

\usepackage{float}

\usepackage{overpic}
\usepackage{subcaption}
\usepackage{graphicx}% Include figure files
\usepackage{dcolumn}% Align table columns on decimal point
\usepackage{bm}% bold math
\usepackage{physics}
\usepackage[colorlinks=true, linkcolor=blue]{hyperref} % 关键包：添加超链接
\usepackage{color}
\usepackage{xcolor}
\usepackage[linesnumbered, ruled, vlined]{algorithm2e}

\begin{document}

\preprint{APS/123-QED}

\title{Remote Adiabatic Controlled-Z Gate between Distant Superconducting Qubits}% Force line breaks with \\

\author{Si-Yu Xu$^{1,2,3}$}
\author{Yi-Rong Jin$^{1}$}
\author{Wen-Gang Zhang$^1$}\email{zhangwg@baqis.ac.cn}
\author{Hai-Feng Yu$^{1,4}$}

\affiliation{$^1$Beijing Academy of Quantum Information Sciences, Beijing 100193, China}
\affiliation{$^2$Institute of Physics, Chinese Academy of Sciences, Beijing 100190, China}
\affiliation{$^3$University of Chinese Academy of Sciences, Beijing 101408, China}
\affiliation{$^4$Hefei National Laboratory, Hefei 230088, China}

\begin{abstract}
Direct entangling gates between spatially separated qubits are a key capability for modular quantum computing. They enable nonlocal quantum circuits to be executed across different processor modules, thereby alleviating single-chip scaling constraints and expanding the effective connectivity of the system. Here, we propose and numerically investigate a remote adiabatic controlled-Z gate between two frequency-tunable superconducting qubits connected by a multi-mode coaxial cable. The cable-mediated interaction enables the gate to be implemented without additional tunable couplers, preserving a simple circuit architecture. For a 30-cm cable, simulations incorporating ten standing-wave modes yield an optimized gate infidelity of $4.71\times10^{-7}$ with a duration of 376.7 ns in the absence of decoherence. The gate maintains high fidelity under qubit-frequency fluctuations and millimeter-scale cable-length variations. Calculations for 50- and 100-cm cable situations demonstrate its potential for even larger cryogenic systems. These results provide a promising approach to extending high-fidelity quantum operations beyond individual chips and toward scalable, interconnected superconducting quantum processors.
% \begin{description}
% Secondary publications and information retrieval purposes.
% You may use the \texttt{description} environment to structure your abstract;
% use the optional argument of the \verb+\item+ command to give the category of each item. 
% \end{description}
\end{abstract}

\maketitle

%\tableofcontents

\textbf{Introductions}---Among the various physical platforms explored for quantum computing\cite{Bluvstein2026,TrappedIonQuantumSimulation,PsiQuantum,PaulSteinacker,Moleculartopological}, superconducting quantum computing has emerged as a promising platform owing to its scalability and tunable interaction capabilities\cite{quantumengineer,Gate-based,jiang2025advancements,9shv-l4cx}. 
Further scaling of superconducting quantum computers is essential for realizing future fault-tolerant processors with large numbers of physical qubits\cite{PhysRevA.86.032324,yoder2506tour,Besedin_2026}. 
However, integrating such large-scale systems onto a single chip remains challenging due to practical limitations, including cryogenic constraints, chip layout and packaging limitations, frequency crowding, and wiring complexity\cite{mohseni2024build,High-density}. 
Distributed quantum computing architectures, which connect multiple quantum processors, therefore offer a viable route toward scalable superconducting quantum systems\cite{jiang2007distributed,ang2024arquin,Fault-Tolerant-Million,main2025distributed}. 
Developing reliable remote interconnection schemes is thus essential for large-scale superconducting quantum computing\cite{zhong2021deterministic,niu2023low,mollenhauer2025high}. 
Among the key challenges is the realization of high-fidelity two-qubit gates between spatially separated qubits.

Recent advances in quantum state transfer (QST) have demonstrated coherent state transmission between spatially separated superconducting qubits through transmission lines\cite{campagne2018deterministic,zhong2019violating,leung2019deterministic,PhysRevLett.125.260502,Qiu2025}. 
However, the QST process requires the target qubit to remain in its ground state, limiting its direct application to deterministic two-qubit gate operations on arbitrary superposition states\cite{cirac1997quantum,korotkov2011flying,kurpiers2018deterministic}. 
Remote entangling gates based on coherent interactions therefore provide a direct approach toward implementing nonlocal two-qubit operations\cite{song2025realization,PhysRevLett.134.020801,ohfuchi2024remote,zhang2025high}. Adiabatic evolution provides a robust approach for superconducting qubit gates by exploiting slowly varying Hamiltonian parameters to maintain instantaneous eigenstates and reduce sensitivity to control errors\cite{Peng2009}. Adiabatic CZ gates exploit state-dependent energy shifts near energy level anti-crossing region to accumulate conditional phases without direct population exchange, suppressing leakage and non-adiabatic transitions while reducing calibration requirements\cite{Martinis2014,Barends2014,Wang2019,HighFidelityHighScalability,Xu2020}. 
Tunable $ZZ$ interactions provide a mechanism for implementing conditional-phase gates through controllable interaction strengths\cite{Collodo2020}. 
Together with optimized control strategies, tunable interactions have enabled high-fidelity adiabatic CZ gates by balancing rapid phase accumulation and adiabatic evolution\cite{HighFidelityHighScalability,Xu2020,Chu2021,Ding2025}. Despite these advances, experimentally demonstrated high-fidelity adiabatic CZ gates have so far been realized primarily between directly coupled qubits on a single chip. 
Meanwhile, remote entangling gates between spatially separated superconducting qubits have relied predominantly on non-adiabatic protocols\cite{PhysRevLett.134.020801,ohfuchi2024remote,zhang2025high}.
In particular, recent experiments have demonstrated high-fidelity remote two-qubit gates through non-adiabatic interaction schemes\cite{song2025realization}, highlighting the feasibility of long-range coherent coupling. However, the feasibility of implementing an adiabatic CZ gate through a remote multi-mode coupling architecture while achieving low leakage and high-fidelity conditional phase accumulation remains unexplored.

In this work, we analyze a remote adiabatic CZ gate between two transmon qubits connected by a coaxial cable. Since tuning the coaxial-cable mode frequencies is technically challenging and would increase the engineering complexity, we instead control the effective interaction strength by tuning the qubit frequencies, thereby avoiding additional tunable elements and preserving a simple circuit architecture. The discrete standing-wave modes of the cable mediate a tunable $ZZ$ interaction, enabling adiabatic accumulation of the conditional phase. Numerical simulations incorporating multiple cable modes and realistic cable parameters yield a gate infidelity below $10^{-6}$ in the absence of decoherence. The gate also retains high fidelity under qubit-frequency fluctuations and millimeter-scale cable-length variations. These results provide a promising route toward modular superconducting quantum processors based on remote adiabatic entangling gates.

\begin{figure}[htbp]
  \centering
  \captionsetup{font={footnotesize}, skip=5pt}
  \begin{subfigure}[b]{0.45\textwidth}
    \centering
    \phantomcaption
    \begin{overpic}[width=\linewidth]{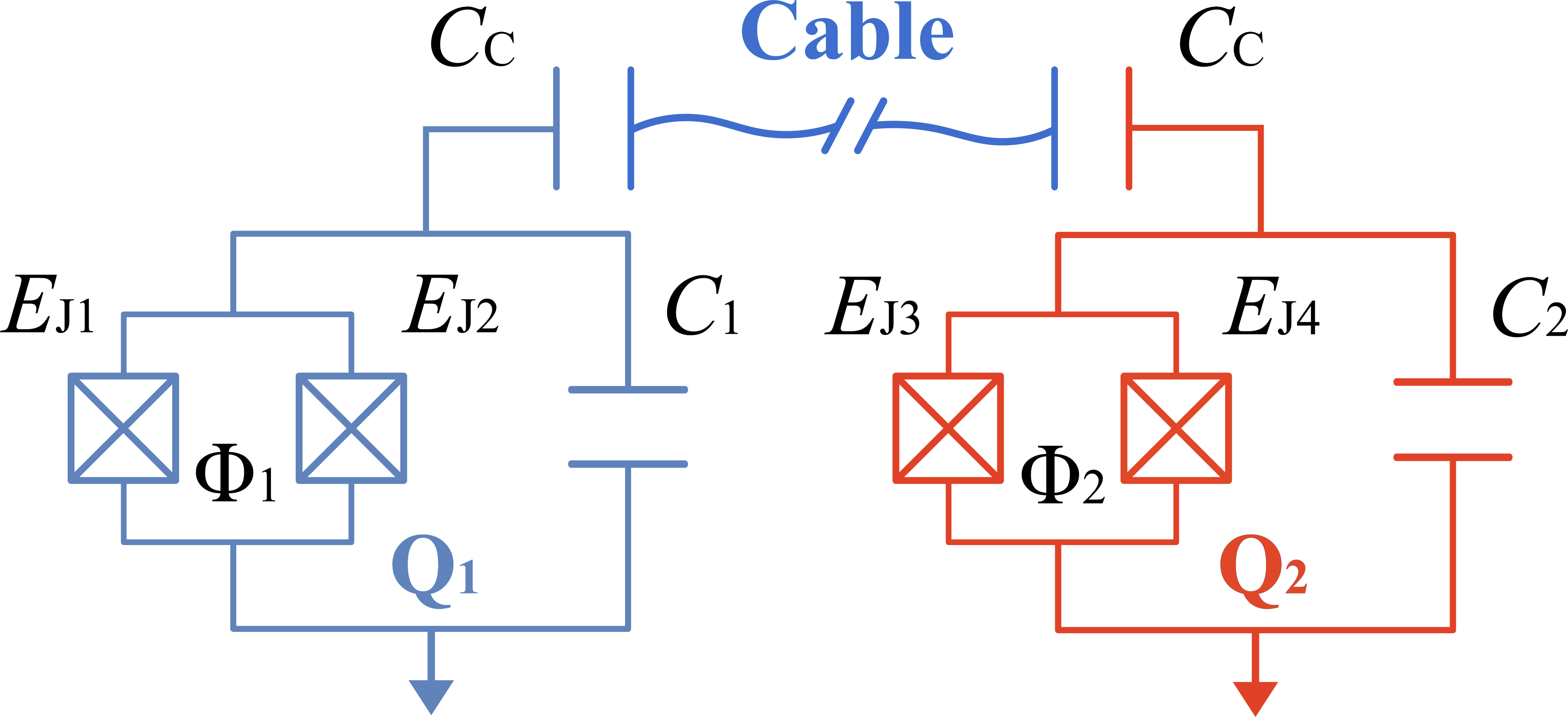}
      \put(0,40){\normalsize (a)} % 左上角标注，数字微调位置
    \end{overpic}
  \end{subfigure}
  \par\vspace{0.3cm}
  \begin{subfigure}[b]{0.45\textwidth}
    \centering
    \phantomcaption
    \begin{overpic}[width=\linewidth]{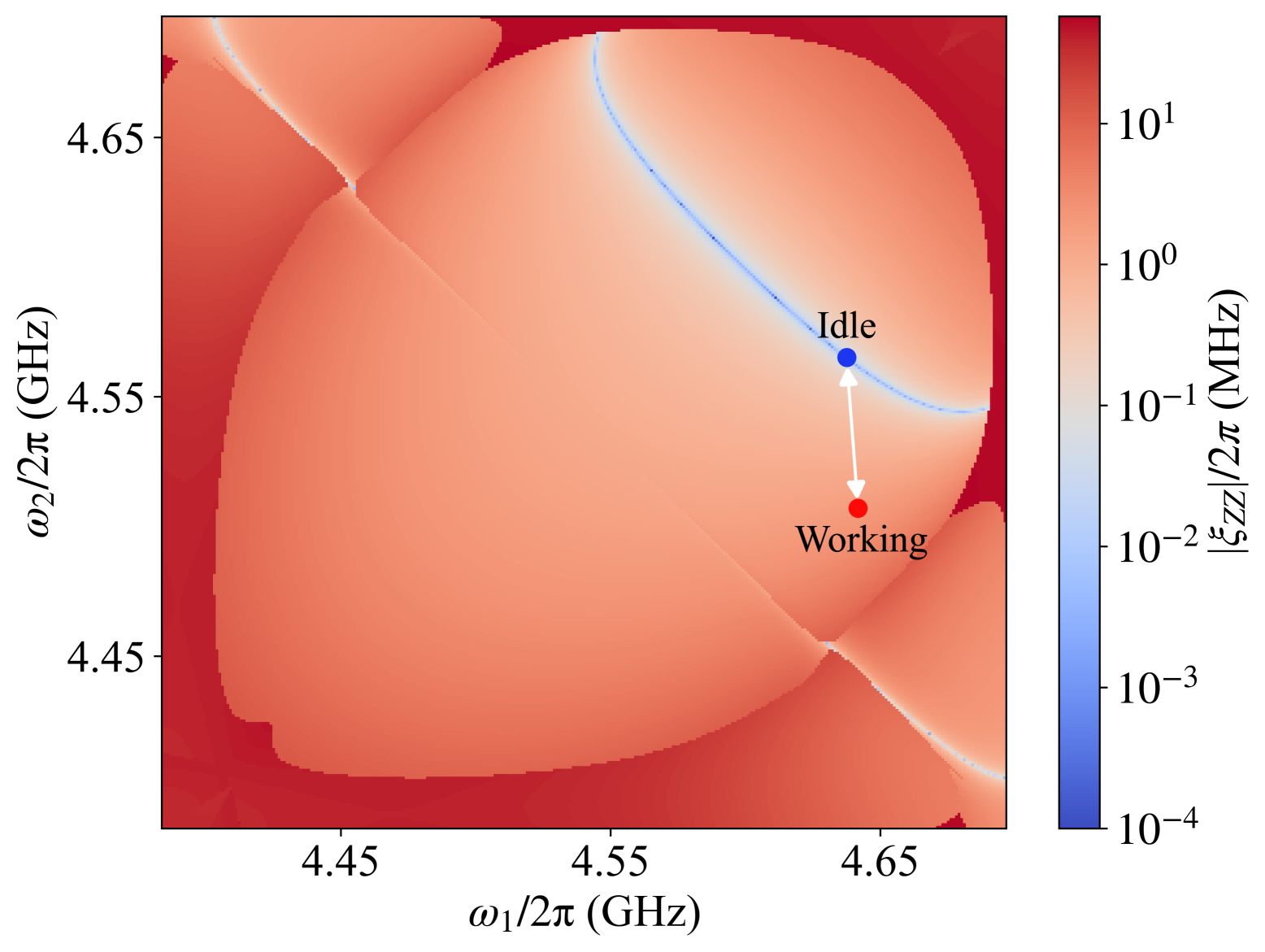}
      \put(0,70){\normalsize (b)} % 左上角标注，数字微调位置
    \end{overpic}
  \end{subfigure}
  \caption{(a) Equivalent circuit diagram of two spatially separated, frequency-tunable transmons ($Q_1$, light blue; $Q_2$, red) coupled remotely via a coaxial cable (dark blue). 
(b) $|\xi_{\text{ZZ}}|$ for a cable length $L_{\text{C}} = 30$ cm. The modes adjacent to the qubit frequencies are $m=14$ and $15$, and the simulation includes modes $m = 10,\dots,19$. The white arrow indicates the frequency tuning path from the idle point (blue dot) to the working point (red dot).}
  \label{fig:1}
\end{figure}

\textbf{System architecture}---As shown in Fig.~\ref{fig:1}(a), the system consists of two spatially separated frequency-tunable transmon qubits, $Q_1$ and $Q_2$, connected by a coaxial cable. Each transmon has a split-junction Superconducting Quantum Interference Device threaded by external magnetic fluxes $\Phi_1$ and $\Phi_2$, allowing frequency tuning. The shunt capacitances $C_1$ and $C_2$ determine the charging energies of the transmons. The Josephson energies are $E_{J1}, E_{J2}$ for $Q_1$ and $E_{J3}, E_{J4}$ for $Q_2$. The coaxial cable is modeled as a multi-mode linear resonator supporting discrete standing-wave modes, with the frequency separation between adjacent modes equal to the free spectral range $\omega_{\text{FSR}}$. Both qubits are coupled to opposite ends of the cable via coupling capacitors $C_\text{c}$.

The system Hamiltonian is given by
\begin{equation}
\begin{aligned}
H &= \sum_{i=1,2}\left[\omega_{i} a_{i}^{\dagger} a_{i}+\frac{\alpha_{i}}{2} a_{i}^{\dagger} a_{i}^{\dagger} a_{i} a_{i}\right]+\sum_{m} m \omega_{\text{FSR}} c_{m}^{\dagger} c_{m} \\
  &\quad +\sum_{m}g_{1, m}\left(a_{1}^{\dagger} c_{m}+c_{m}^{\dagger}a_{1}\right)\\
  &\quad +\sum_{m}(-1)^{m} g_{2, m}\left(a_{2}^{\dagger} c_{m}+c_{m}^{\dagger} a_{2}\right),
\end{aligned}
\label{eq1}
\end{equation}
where $a_i^\dagger$ ($a_i$) and $c_m^\dagger$ ($c_m$) are the creation (annihilation) operators for the qubits and the $m$-th cable mode, respectively. For each qubit $i$, $\omega_i$ is the transition frequency between $|0\rangle$ and $|1\rangle$, and $\alpha_i$ is its anharmonicity. $g_{i,m}$ is the transverse coupling strength between the $i$-th qubit and the $m$-th cable mode. The factor $(-1)^m$ arises from the alternating voltage polarity of adjacent standing-wave modes at the two ends of the cable. We define the ZZ interaction strength as
\begin{equation}
\xi_{\text{ZZ}} = E_{11} + E_{00} - E_{10} - E_{01},
\label{eq2}
\end{equation}
where $E_{ij}$ denotes the eigenenergy of the dressed eigenstate corresponding to the bare state $|ij\rangle\otimes|\mathbf{0}\rangle_c$, and $|\mathbf{0}\rangle_c$ denotes the joint vacuum state of all cable modes.

\textbf{Adiabatic gate protocol---}The qubit transition frequencies $\omega_i/2\pi$ ($i=1,2$) are tuned within the range of $4.3$ to $4.7$ GHz, with shunt capacitances $C_1=C_2=90\text{fF}$ and anharmonicities $\alpha_1/2\pi=\alpha_2/2\pi=-200\,\text{MHz}$. 
We use a 30~cm aluminum coaxial cable with polytetrafluoroethylene insulation, with a relative permittivity $\epsilon_r=2.55$. 
The cable capacitance and free spectral range are approximately $31.9\,\text{pF}$ and $2\pi\times313.1\,\text{MHz}$, respectively. 
The modes $m=14$ and $m=15$ reside near the qubit working frequencies.
Based on numerical optimization, the qubit-cable coupling capacitance is chosen as $C_\text{c}=20\text{fF}$ to enable high-contrast ZZ-interaction [see Supplemental Material (SM) for details].

Fig.~\ref{fig:1}(b) shows the ZZ interaction strength $|\xi_{\text{ZZ}}|$ over a frequency range of one FSR, i.e., the interval between modes $m=14$ and $m=15$. We choose the minimum-$|\xi_{\text{ZZ}}|$ point as the idle point, with $|\xi_{\text{ZZ}}| \ll 2\pi\times10$~kHz.
Candidate working points are selected from regions with large $|\xi_{\text{ZZ}}|$, which helps shorten the gate duration.
Furthermore, the gate trajectory is chosen not to pass through the energy level anti-crossing regions, thereby suppressing possible non-adiabatic leakage.
Due to the exchange symmetry of the system, the $|\xi_{\text{ZZ}}|$ distribution is symmetric about the diagonal in Fig.~\ref{fig:1}(b).
Hence, we restrict our analysis to the region below that diagonal.

\begin{figure}[htbp]
  \centering
  \captionsetup{font={footnotesize}, skip=5pt}
  \begin{subfigure}[b]{0.45\textwidth}
    \centering
    \phantomcaption
    \begin{overpic}[width=\linewidth]{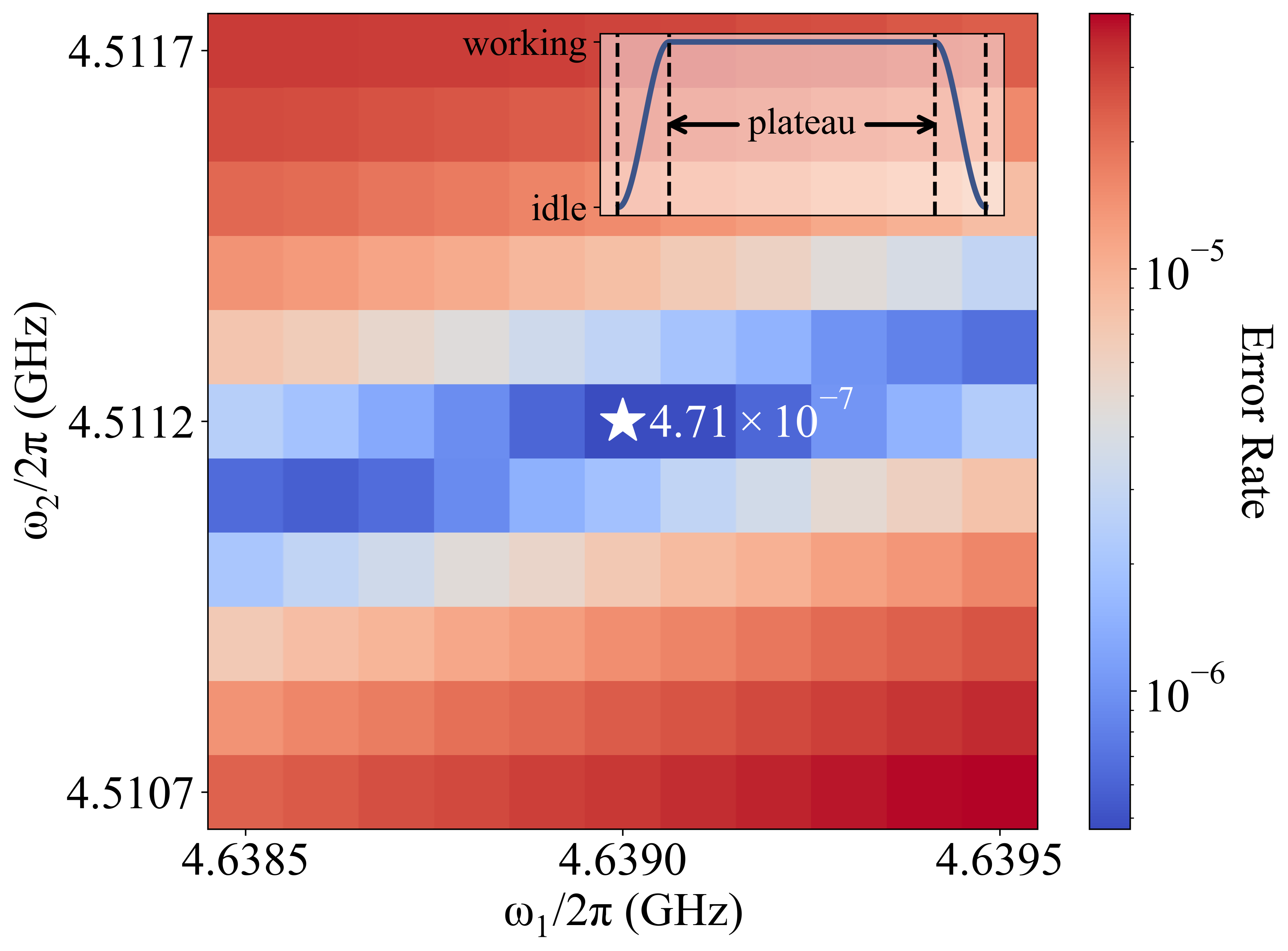}
      \put(-1,69){\normalsize (a)} % 左上角标注，数字微调位置
    \end{overpic}
  \end{subfigure}
   \par\vspace{0.25cm}
  \begin{subfigure}[b]{0.43\textwidth}
    \centering
    \phantomcaption
    \begin{overpic}[width=\linewidth]{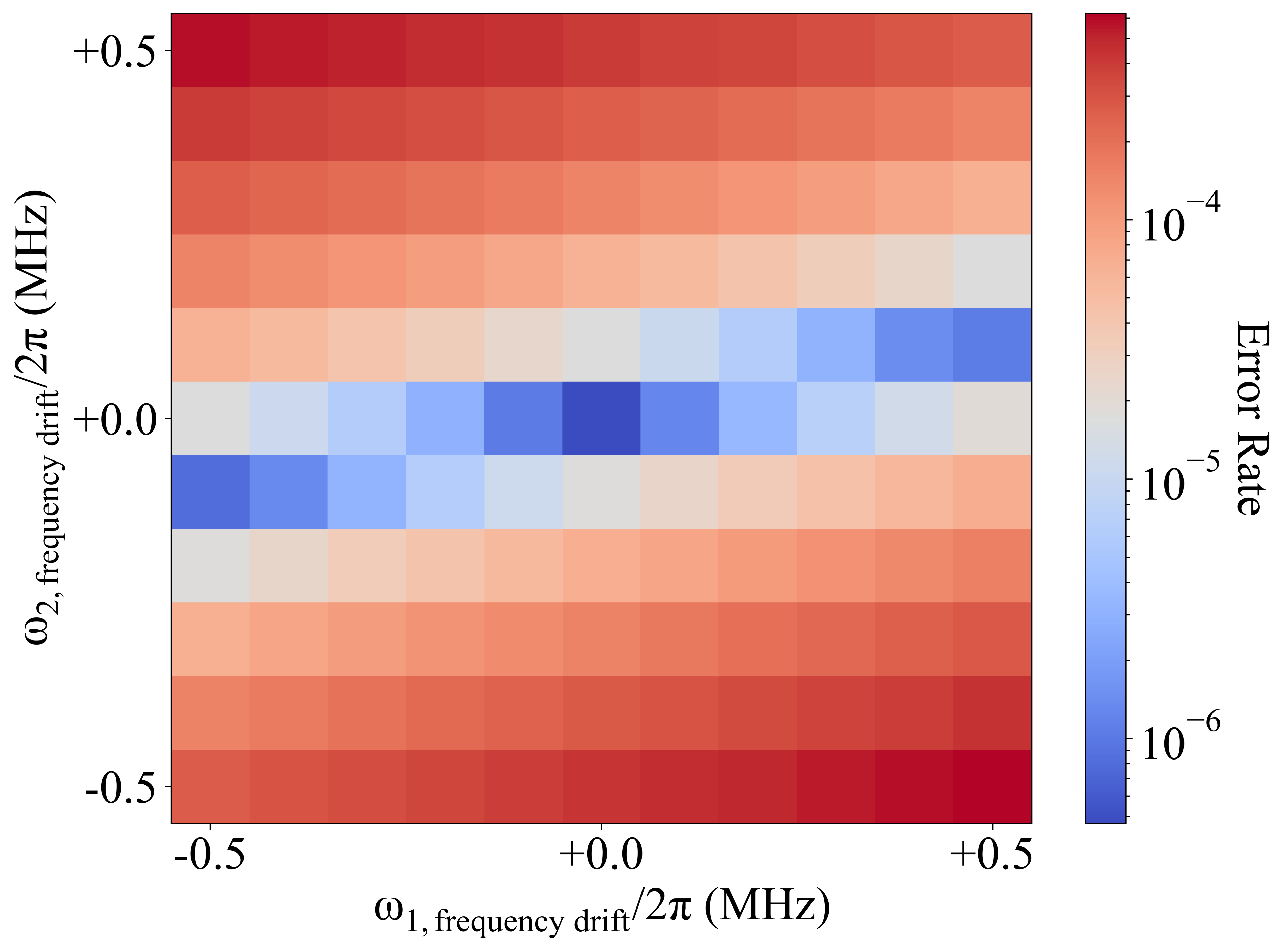}
      \put(-2,69){\normalsize (b)} % 左上角标注，数字微调位置
    \end{overpic}
  \end{subfigure}
  \par\vspace{0.25cm}
  \begin{subfigure}[b]{0.45\textwidth}
    \centering
    \begin{overpic}[width=\linewidth]{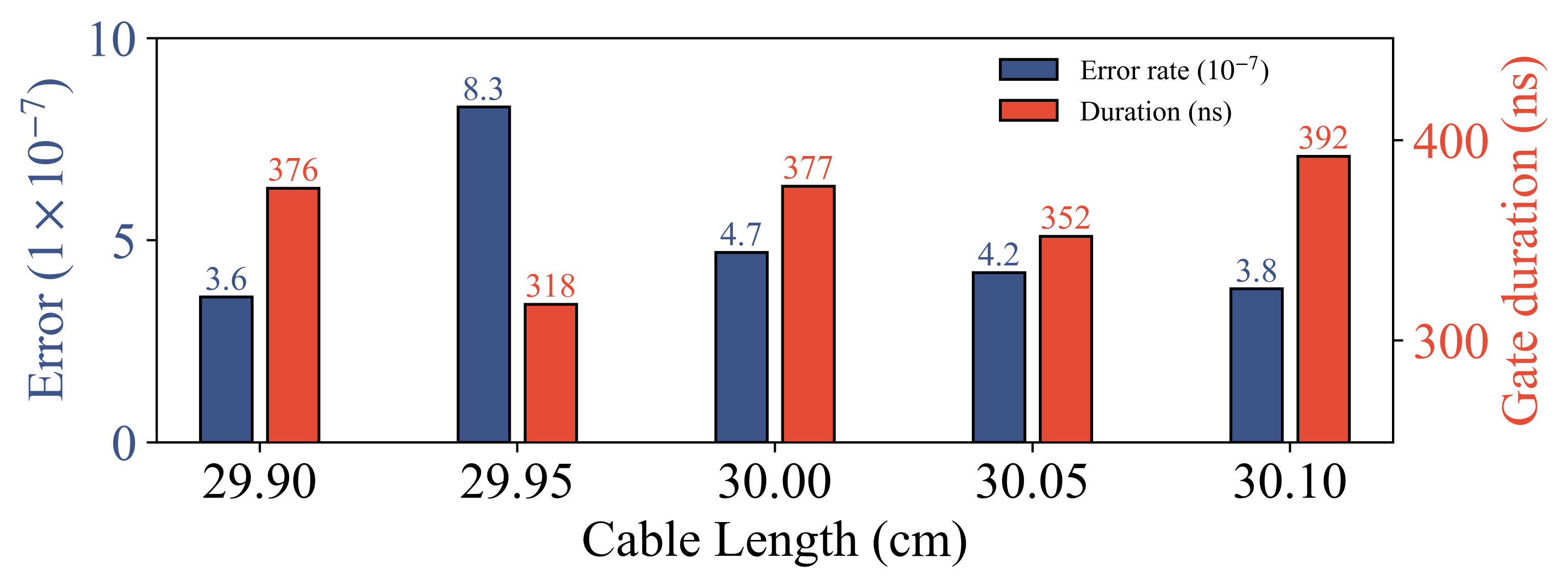}
      \put(-1,37){\normalsize (c)} % 左上角标注，数字微调位置
    \end{overpic}
  \end{subfigure}
  \caption{ (a) Optimized gate infidelity as a function of the two-qubit working frequencies. The gate duration is optimized independently at each frequency point. The minimum infidelity of $4.71\times10^{-7}$ is obtained at $(4.6390,4.5112)~\mathrm{GHz}$. The inset shows the control pulse. (b) Gate infidelity under qubit-frequency fluctuations of up to $\pm0.5~\mathrm{MHz}$, with the pulse amplitudes and gate duration fixed and only the single-qubit phases re-calibrated. (c) Optimized gate infidelity (left axis, blue bars) and gate duration (right axis, red bars) as functions of the cable length, sampled over $30.00\pm0.10~\mathrm{cm}$ at intervals of $0.5~\mathrm{mm}$. The qubit working point is re-optimized at each cable length.}
  \label{fig:2}
\end{figure}

The gate is implemented by tuning the qubit frequencies from the idle point to the working point, holding them there to accumulate the required conditional phase, and subsequently returning them to the idle point, as indicated by the white arrow in Fig. ~\ref{fig:1}(b).
We set the duration of the rising and falling cosine-shaped edges to 50 ns, and then scan the plateau duration to maximize the gate fidelity, as illustrated in the inset of Fig.~\ref{fig:2}(a).
We first calculate the full-system evolution operator generated by the time-dependent Hamiltonian over the complete control pulse,
\begin{equation}
    U(t)=\mathcal{T}\exp\left[-i\int_{0}^{t}H(\tau)d\tau\right],
\end{equation}
where $\mathcal{T}$ denotes time ordering. We then project $U(t)$ onto the dressed computational basis at the idle point $\{ |\widetilde{00}\rangle,
|\widetilde{01}\rangle,
|\widetilde{10}\rangle,
|\widetilde{11}\rangle\}$.
The matrix elements of the resulting $4\times4$ projected evolution matrix are:
\begin{equation}
    (U_{proj})_{ij,kl}=\langle\widetilde{ij}|
U(t)
|\widetilde{kl}\rangle,
\qquad
i,j,k,l\in \{0,1\}.
\end{equation}
The rows and columns of $U_{\mathrm{proj}}$ follow the basis order
$|\widetilde{00}\rangle,|\widetilde{01}\rangle,
|\widetilde{10}\rangle,$ and $|\widetilde{11}\rangle$.
Since a small amount of population may remain outside the computational subspace at the end of the gate, $U_{\mathrm{proj}}$ is not necessarily unitary.
The single-qubit dynamical phases accumulated during the pulse are compensated by virtual-$Z$ rotations. The resulting phase-corrected matrix is then denoted by $U_{\mathrm{sim}}$.
The gate fidelity is defined as
\begin{equation}
F = \frac{|\mathrm{Tr}(U_\mathrm{ideal}^\dagger U_\mathrm{sim})|^2+\mathrm{Tr}(U_\mathrm{sim}^\dagger U_\mathrm{sim})}{d(d+1)},
\label{eq3}
\end{equation}
where $d=4$ denotes the dimension of the two-qubit computational subspace, and $U_\mathrm{ideal}$ is the unitary matrix of the ideal CZ gate.

Using the above protocol, we optimize the gate fidelity over the working point and plateau duration. As shown in Fig.~\ref{fig:2}(a), a gate infidelity of $4.71\times10^{-7}$ is achieved at $(4.6390, 4.5112)$~GHz, with a total gate duration of $376.7~ns$.

\textbf{Gate robustness analysis}---To account for qubit-frequency drift caused by environmental noise and cable-length deviations arising from fabrication tolerances, we evaluate the robustness of the gate against both effects.

To evaluate the robustness against qubit-frequency fluctuations, we introduce frequency offsets of up to $\pm0.5~\mathrm{MHz}$ around the optimized working point while keeping the pulse amplitudes and gate duration fixed. As shown in Fig.~\ref{fig:2}(b), after re-calibrating only the single-qubit phases, the gate infidelity remains below $7\times10^{-4}$ throughout the investigated range. Since single-qubit phases can be calibrated relatively efficiently in experiments, these results indicate that high-fidelity remote CZ gates can be maintained under moderate qubit-frequency fluctuations without re-optimizing the two-qubit gate pulse.

For cable length errors, fabrication tolerances on the order of millimeters are unavoidable. We therefore examine the gate fidelity for cable lengths sampled at 0.5 mm intervals within the 30~cm $\pm$~1~mm range. As shown in Fig.~\ref{fig:2}(c), after re-optimizing the working point at each cable length, we identify a high-fidelity operating regime for every tested length, demonstrating tolerance to fabrication-induced length variations with re-calibration.

For cable lengths of 50 and 100 cm, the optimized gate infidelities reach $2.22\times10^{-6}$ and $8.28\times10^{-6}$, respectively. Under qubit-frequency fluctuations of up to $\pm0.5$ MHz, with the pulse parameters fixed and only the single-qubit phases re-calibrated, the maximum infidelities increase to $1.53\times10^{-3}$ and $1.95\times10^{-2}$, respectively (see the SM for details). These results indicating its applicability to larger cryogenic systems.

\textbf{The influence of the included cable modes}---In numerical simulations, including more cable modes yields a more accurate description of the realistic multi-mode behavior. However, retaining too many modes leads to a rapidly growing Hilbert space dimension and prohibitively long computation times. Therefore, an appropriate truncation of the cable modes is essential for efficient and accurate simulation.
\begin{figure}[htbp]
  \centering
  \captionsetup{font={footnotesize}, skip=5pt}
  \begin{subfigure}[b]{0.45\textwidth}
    \centering
    \phantomcaption
    \includegraphics[width=\linewidth]{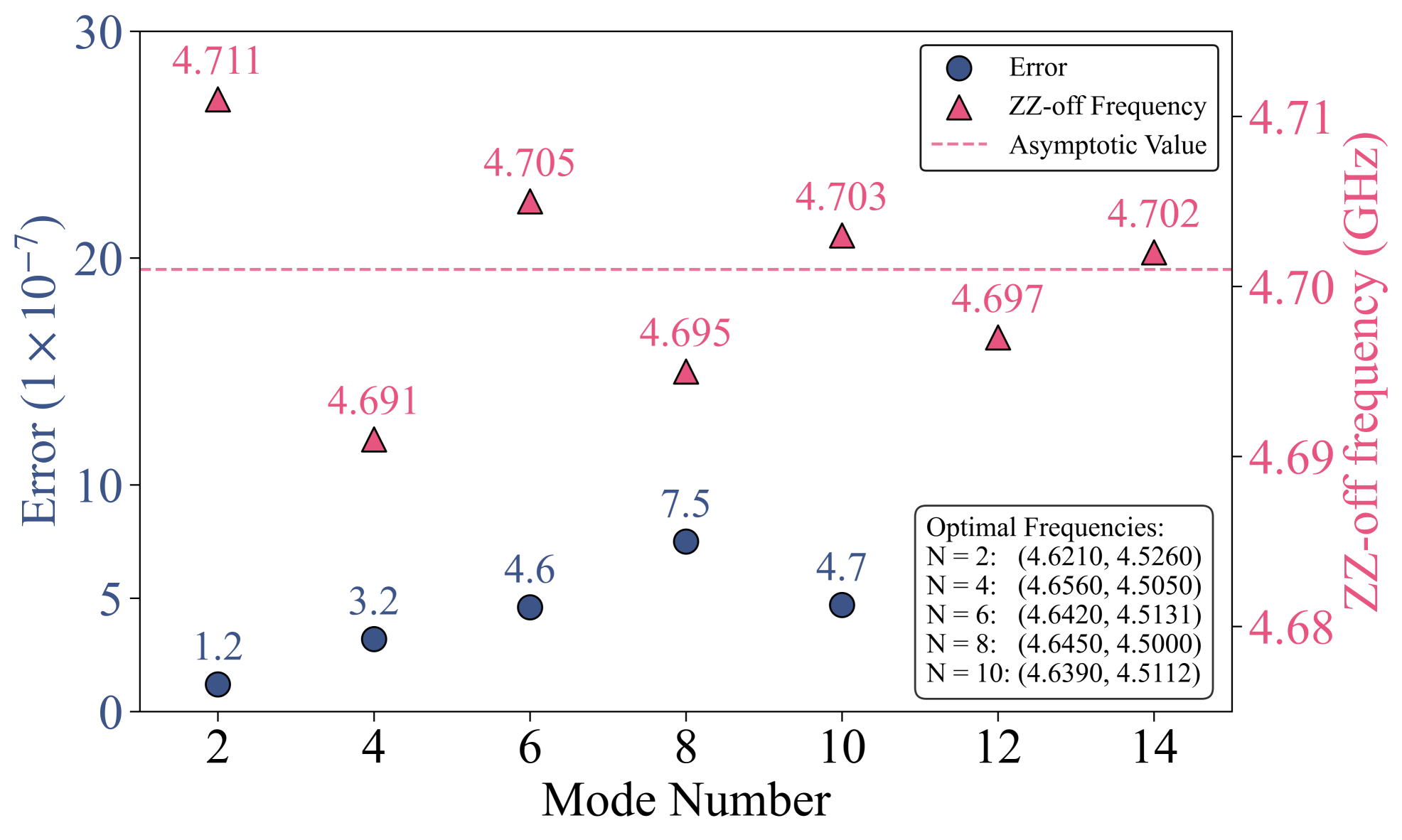}
  \end{subfigure}
  \caption{The ZZ-off frequency (pink triangles) at $\omega_1=\omega_2$ converges as the number of cable modes increases. Working points with gate infidelity below $10^{-6}$ are found for mode numbers of 2, 4, 6, 8, and 10, while incorporating more cable modes increases the computation cost. The inset shows the optimal working points for different mode numbers.}
  \label{fig:3}
\end{figure}

Fig.~\ref{fig:3} shows the ZZ-off frequency (pink triangles) when $\omega_1=\omega_2$ and the optimized gate errors (blue dots) as a function of the number of included cable modes. The ZZ-off frequency converges as the number of cable modes increases, indicating that modes with frequencies farther from the qubit frequencies contribute less to the ZZ coupling. The maximum gate infidelity remains below $10^{-6}$ when 2 to 10 modes are included, with the optimal working points for different mode numbers shown in the inset at the bottom-right of Fig.~\ref{fig:3}. These results support the experimental feasibility of the scheme.

However, incorporating more cable modes increases the computation time. For the scan in Fig.~\ref{fig:2}(a), the total computation time is 6 hours for 10 modes, 15 hours for 12 modes, and 36 hours for 14 modes. Considering the computational cost and the convergence of the results in Fig.~\ref{fig:3}, we use 10 modes in all subsequent simulations.

\begin{figure}[htbp]
  \centering
  \captionsetup{font={footnotesize}, skip=-5pt}
  \begin{subfigure}[b]{0.45\textwidth}
    \centering
    \begin{overpic}[width=\linewidth]{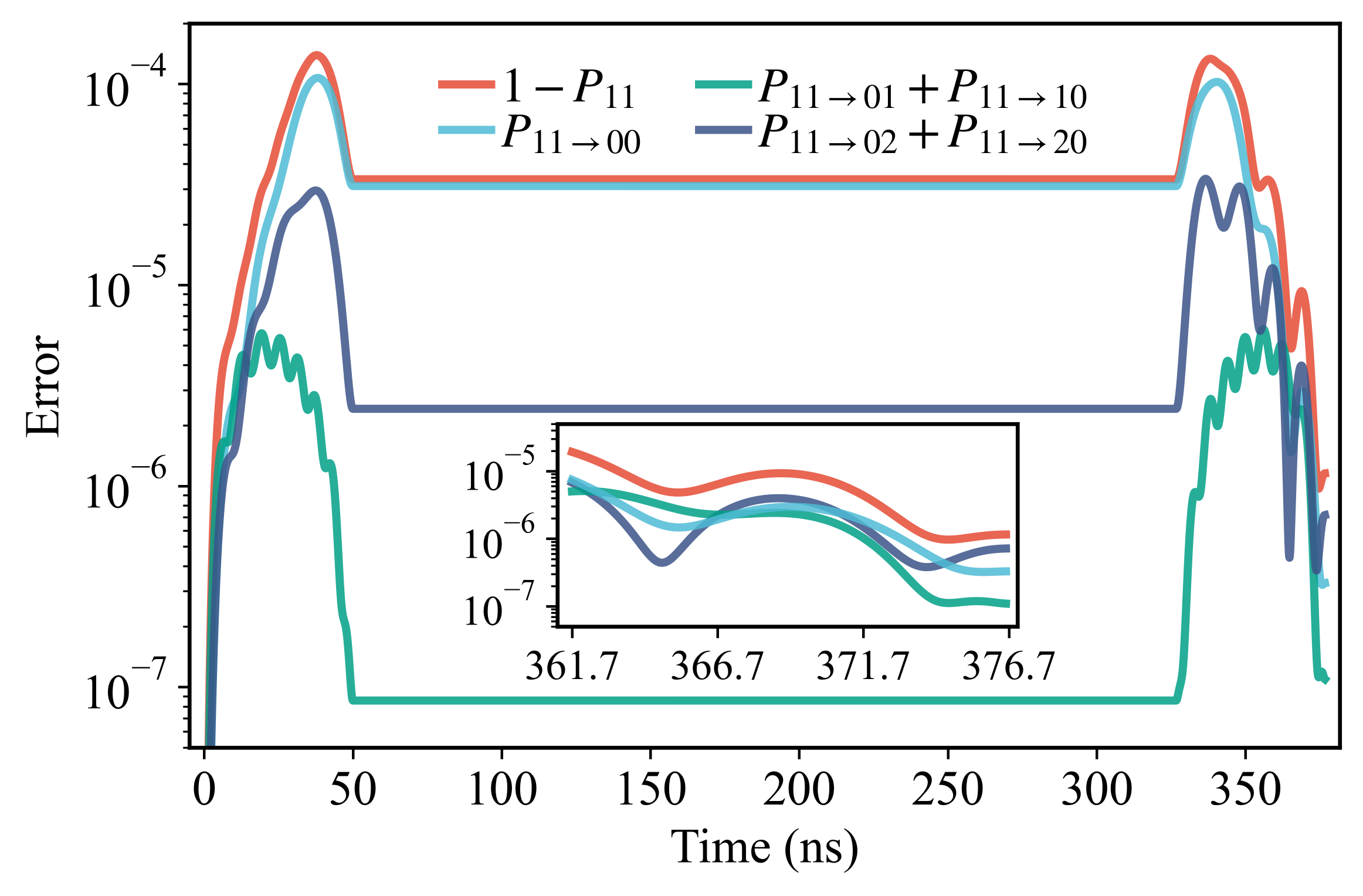}
      \put(0,61){\normalsize (a)} % 左上角标注，数字微调位置
    \end{overpic}
    \label{fig:4a}
  \end{subfigure}
  \par\vspace{-0.35cm}
  \begin{subfigure}[b]{0.45\textwidth}
    \centering
    \begin{overpic}[width=\linewidth]{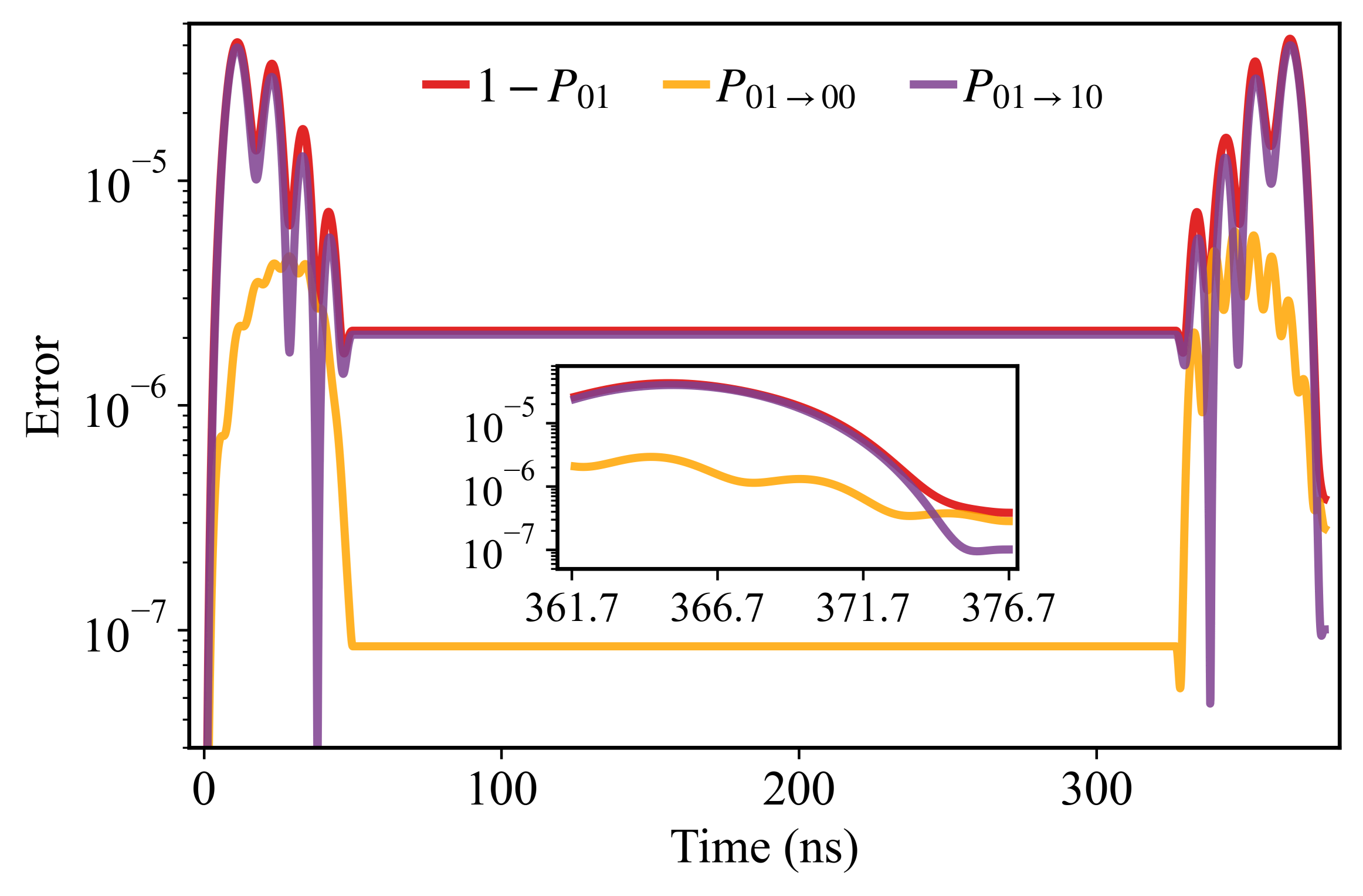}
      \put(0,61){\normalsize (b)} % 左上角标注，数字微调位置
    \end{overpic}
    \label{fig:4b}
  \end{subfigure}
  \caption{(a) Time evolution of population-transfer error for the $|\widetilde{11}\rangle$ initial state. $\varepsilon_{11}=1-P_{11}$ (red) is decomposed into leakage to non-computational qubit states $P_{02}+P_{20}$ (dark blue), transfer to $|\widetilde{01}\rangle/|\widetilde{10}\rangle$ with one cable photon $P_{01}+P_{10}$ (teal), and transfer to $|\widetilde{00}\rangle$ with two cable photons $P_{00}$ (light blue). The inset magnifies the final 15 ns. (b) Time evolution of population-transfer error for the $|\widetilde{01}\rangle$ initial state. $\varepsilon_{01}=1-P_{01}$ (red) consists of leakage to $|\widetilde{00}\rangle$ with one cable photon $P_{00}$ (yellow) and swap error $P_{10}$ (purple). The inset magnifies the final 15 ns.}
  \label{fig:4}
\end{figure}

\textbf{Error analysis}---We next analyze the population-transfer errors during the gate operation at the optimized working point for the 30-cm cable. These errors include both leakage outside the computational subspace and unwanted population transfer within it.
For an initial computational state $|\widetilde{ij}\rangle_0$, we define the total population-transfer error as
\begin{equation}
\varepsilon_{ij} = 1 - P_{ij}, \qquad  P_{ij}=|\langle \widetilde{ij}|_t U(t)|\widetilde{ij}\rangle_0 |^{2}
\label{eq:eps},
\end{equation}
where $|\widetilde{ij}\rangle_t$ is the instantaneous eigenstate corresponding to the bare computational state $|ij\rangle$, and $U(t)$ is the time evolution operator of the full qubit-cable system.
Although transient population transfer occurs during the pulse, most of the population returns to the target state as the qubits are brought back to the idle point.
We therefore focus on the residual error at the end of the gate.
Detailed definitions and a mode-resolved decomposition are provided in the SM.

We first consider the system initialized in $|\widetilde{11}\rangle$. As shown in Fig. 4(a), the population transferred out of the target instantaneous state can be divided into three categories: transitions to the non-computational qubit states $|\widetilde{02}\rangle$ and $|\widetilde{20}\rangle$ ($P_{11\rightarrow02}+P_{11\rightarrow20}$); transitions to states associated with $|01\rangle$ or $|10\rangle$ and one cable photon ($P_{11\rightarrow01}+P_{11\rightarrow10}$); and transitions to states associated with $|00\rangle$ and two cable photons ($P_{11\rightarrow00}$). At the end of the gate, the total population-transfer error is $\varepsilon_{11}=1.16\times10^{-6}$. Among the three categories, leakage to the non-computational qubit subspace contributes approximately $7.17\times10^{-7}$, two-photon excitation of the cable contributes $3.29\times10^{-7}$, and the combined single-photon channels contribute $1.09\times10^{-7}$.

For the $|\widetilde{01}\rangle$ input state, as shown in Fig. 4(b), the population-transfer error contains two contributions: the transfer of the qubit excitation to the cable ($P_{01\rightarrow00}$), and the population-swap error from $|\widetilde{01}\rangle$ to $|\widetilde{10}\rangle$ ($P_{01\rightarrow10}$).
At the end of the gate, the total error is \(\varepsilon_{01}=3.84\times10^{-7}\), of which the photon leakage error is $P_{01\rightarrow00}=2.83\times10^{-7}$ and the population-swap error is $P_{01\rightarrow10}=1.01\times10^{-7}$.
Owing to the exchange symmetry of the system, the $|\widetilde{10}\rangle$ input state exhibits a similar population evolution, with a final error of $\varepsilon_{10}=3.76\times10^{-7}$; its detailed decomposition is presented in the SM.

For the $|\widetilde{00}\rangle$ input state, excitation-number conservation prevents the global ground state from coupling to any excited state, and therefore $\varepsilon_{00}=0$ throughout the gate operation.

Overall, the residual population-transfer errors at the end of the gate remain at or below \(1.16\times10^{-6}\) for all computational input states. These results show that most of the transient population returns to the target instantaneous states at the end of the pulse, leaving only small residual non-adiabatic errors.

\textbf{Discussions}---In summary, we have numerically demonstrated a remote adiabatic CZ gate between two transmon qubits connected by a coaxial cable. By adiabatically tuning the qubit frequencies, the $ZZ$ interaction mediated by the cable modes accumulates a conditional phase of $\pi$ over time. At the optimal working point for a 30 cm cable, we achieve a gate infidelity below $10^{-6}$. We have verified the robustness of the gate against $\pm0.5$ MHz frequency fluctuations and $\pm1$ mm cable length variations.
High-fidelity remote gates can also be achieved with 50-cm and 100-cm cables, supporting the applicability of this approach to larger cryogenic systems.
Our results thus demonstrate a practical path toward modular superconducting quantum processors, offering simple, high-fidelity remote two-qubit gates over centimeter-to-meter distances. 
\section*{Acknowledgements}
We thank Dr. Pei Liu, Dr. Yulong Feng, Dr. Zhaobo Wang, and Dr. Huikai Xu for their valuable discussions and suggestions.  This work was supported by the National Natural Science Foundation of China (No. 12404557, No. 92365206 and No. 92565301), and the Quantum Science and Technology-National Science and Technology Major Project(No. 2021ZD0301802, No. 2024ZD0301500 and No.2023ZD0300200).

%
% ****** Main-text bibliography ******
%

% ============================================================
%                     SUPPLEMENTAL MATERIAL
% ============================================================

\clearpage
\onecolumngrid

% 补充材料图表编号加S前缀
\makeatletter
\setcounter{figure}{0}
\setcounter{table}{0}
\renewcommand{\fnum@figure}{FIG. \thefigure}
\renewcommand{\thefigure}{S\arabic{figure}}
\renewcommand{\thesubfigure}{\alph{subfigure}}
\renewcommand{\p@subfigure}{S\thefigure}
\renewcommand{\fnum@table}{TABLE S\thetable}
\makeatother

\captionsetup{labelfont=rm, labelsep=period}

\section{Details of ZZ interaction}

\begin{figure}[htbp]
  \centering
  % \captionsetup{font={footnotesize}, skip=-5pt}
  % 子图(a)
  \begin{subfigure}[b]{0.32\textwidth}
    \centering
    % [width=...]控制图宽，(10,85)是坐标：左下角原点，x右/y上
    \begin{overpic}[width=\linewidth]{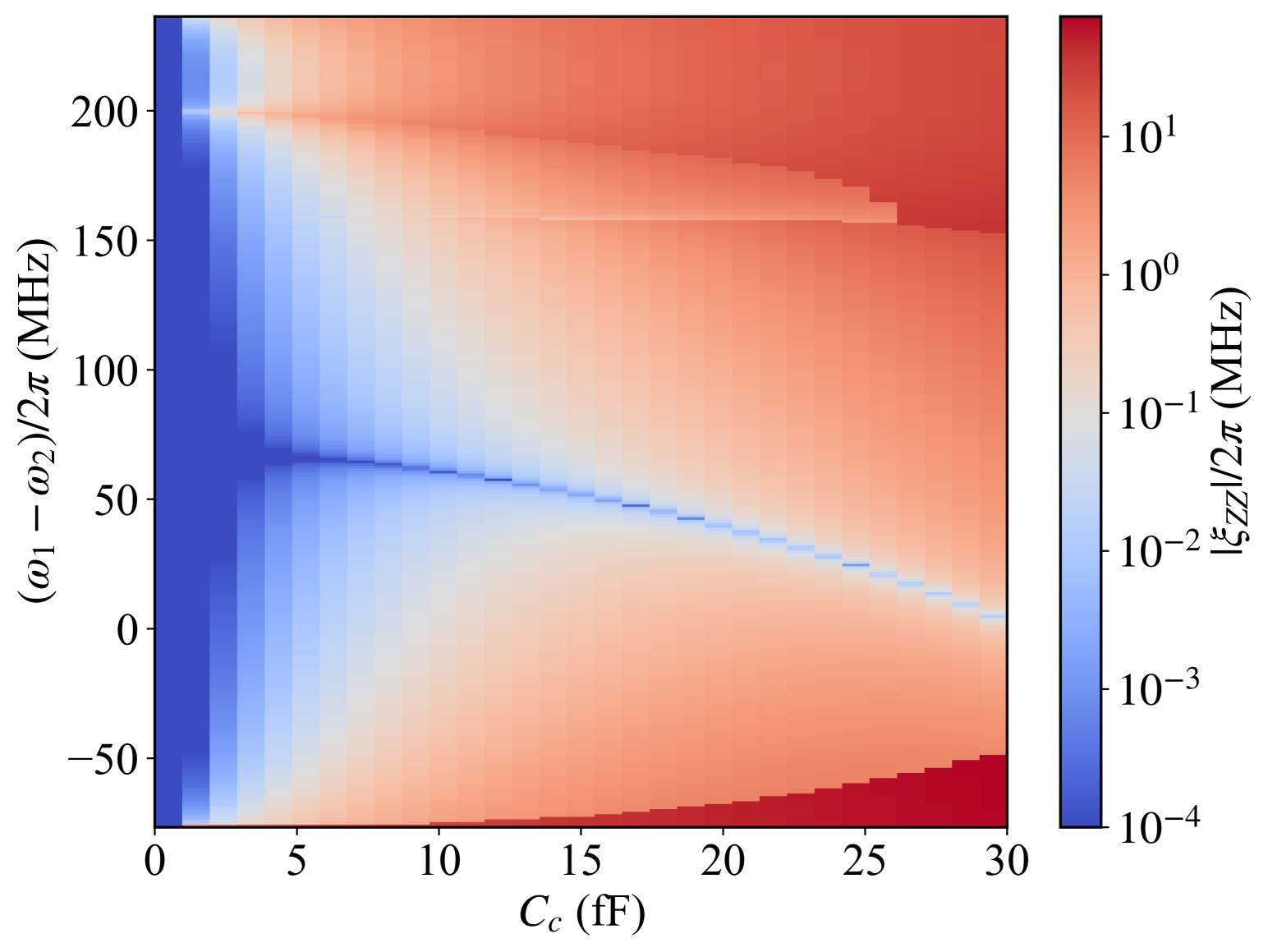}
      \put(-3,68){\normalsize (a)}
    \end{overpic}
    \label{fig:S1a}
  \end{subfigure}
  \hspace{0.1cm}
  % 子图(b)
  \begin{subfigure}[b]{0.32\textwidth}
    \centering
    \begin{overpic}[width=\linewidth]{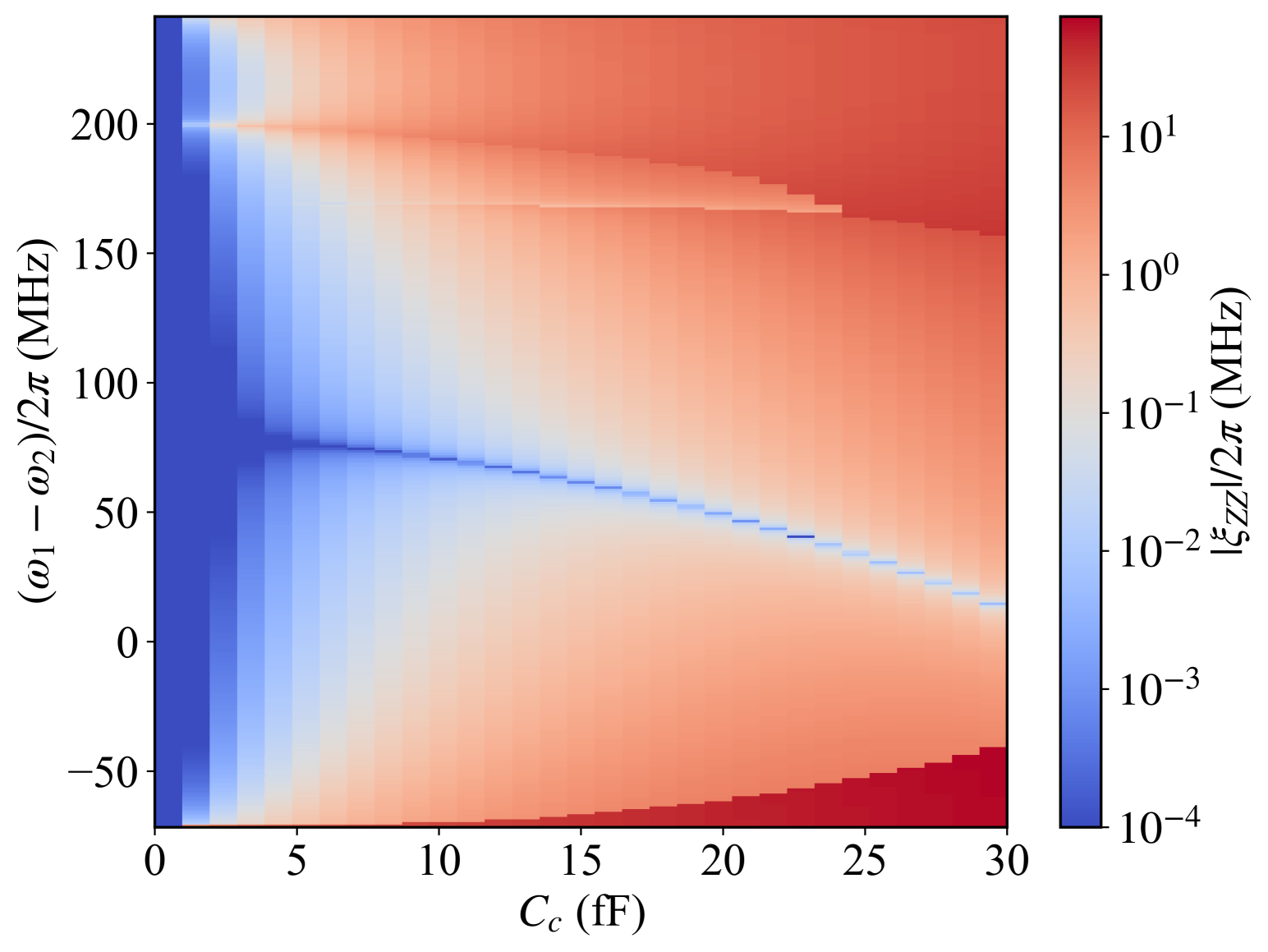}
      \put(-3,68){\normalsize (b)}
    \end{overpic}
    \label{fig:S1b}
  \end{subfigure}
  \hspace{0.1cm}
  % 子图(c)
  \begin{subfigure}[b]{0.32\textwidth}
    \centering
    \begin{overpic}[width=\linewidth]{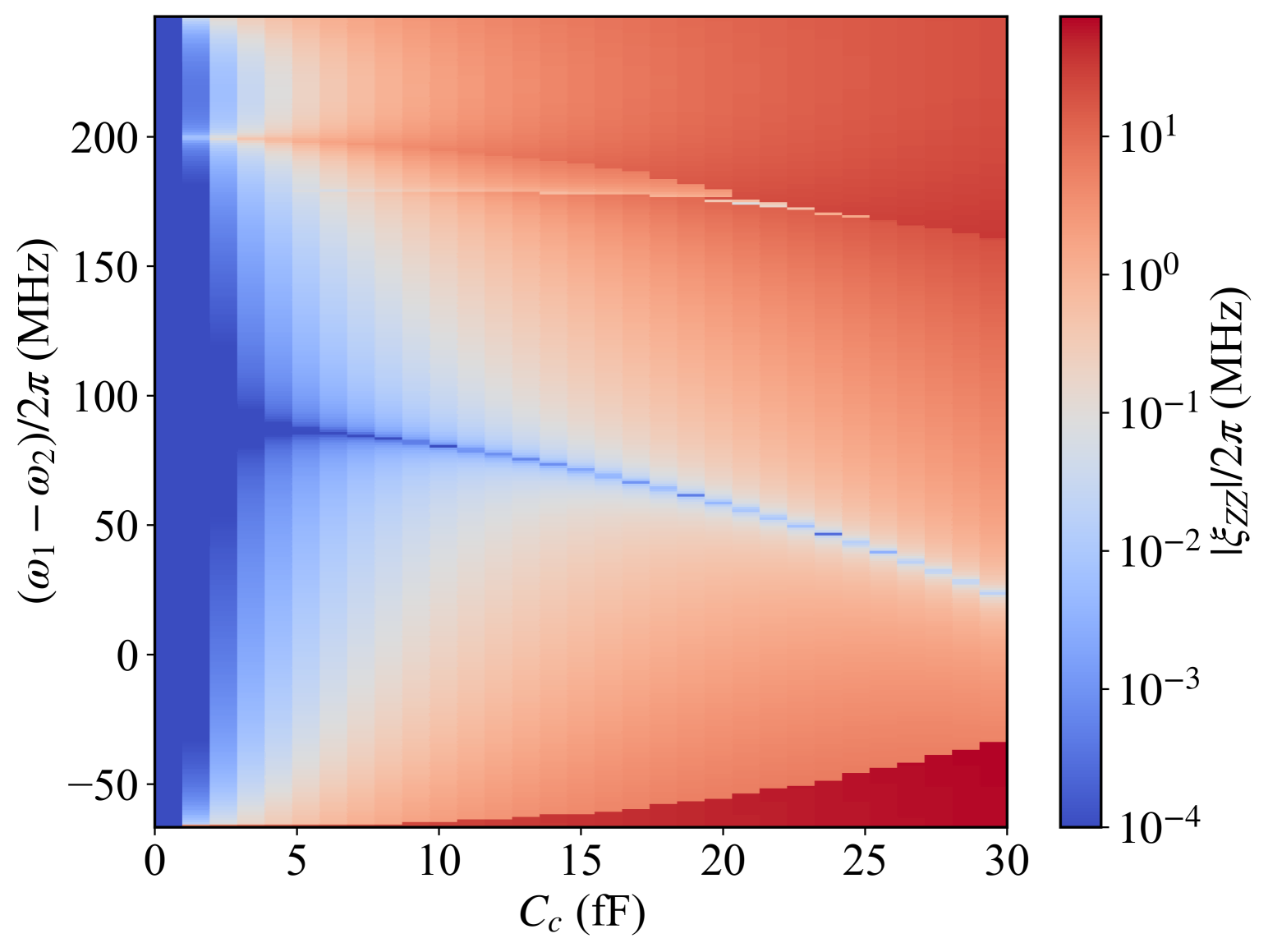}
      \put(-3,68){\normalsize (c)}
    \end{overpic}
    \label{fig:S1c}
  \end{subfigure}
  \hspace{0.1cm}
  % 子图(d)
  \begin{subfigure}[b]{0.32\textwidth}
    \centering
    \begin{overpic}[width=\linewidth]{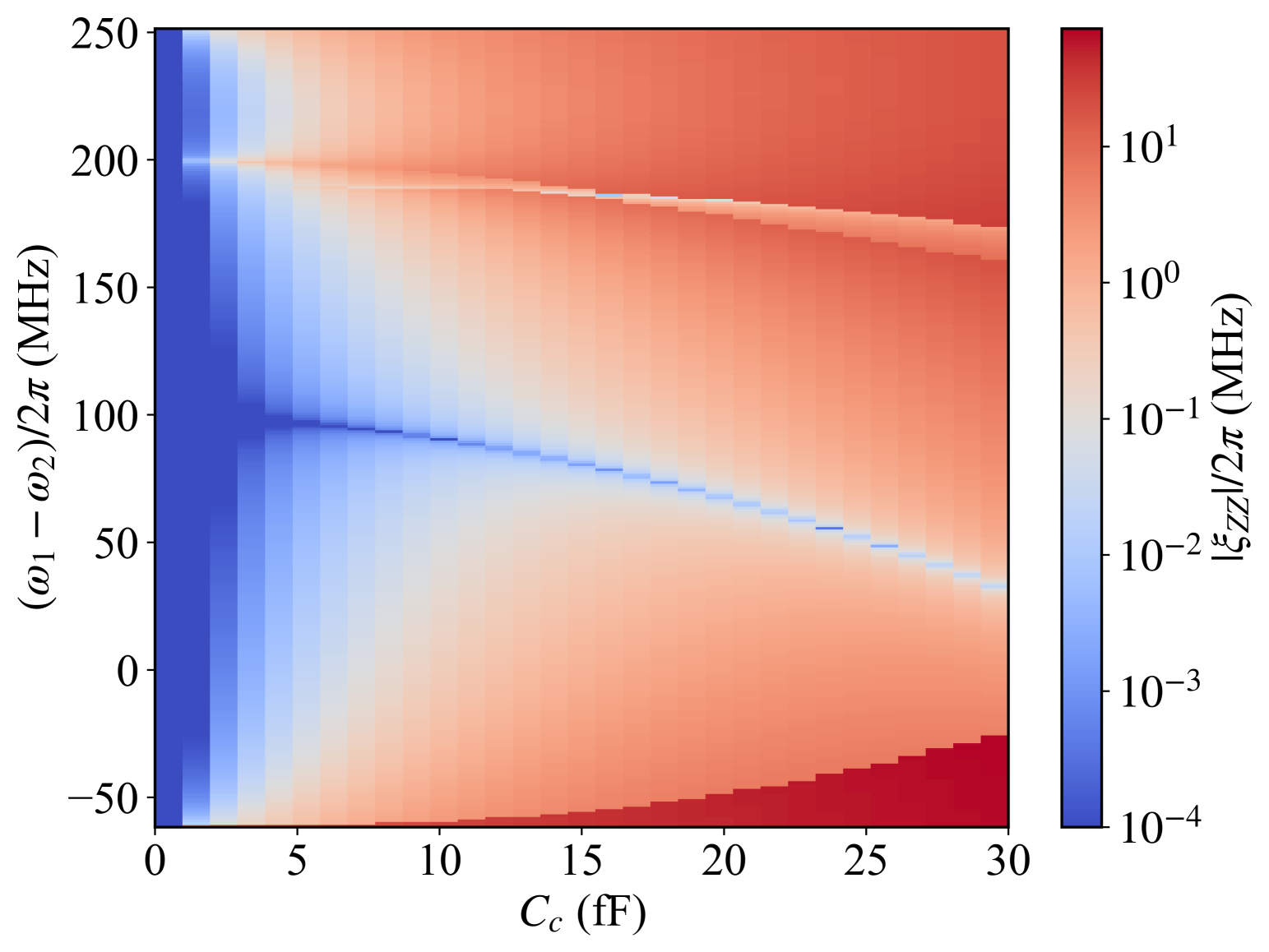}
      \put(-3,68){\normalsize (d)}
    \end{overpic}
    \label{fig:S1d}
  \end{subfigure}
  \hspace{0.1cm}
  % 子图(e)
  \begin{subfigure}[b]{0.32\textwidth}
    \centering
    \begin{overpic}[width=\linewidth]{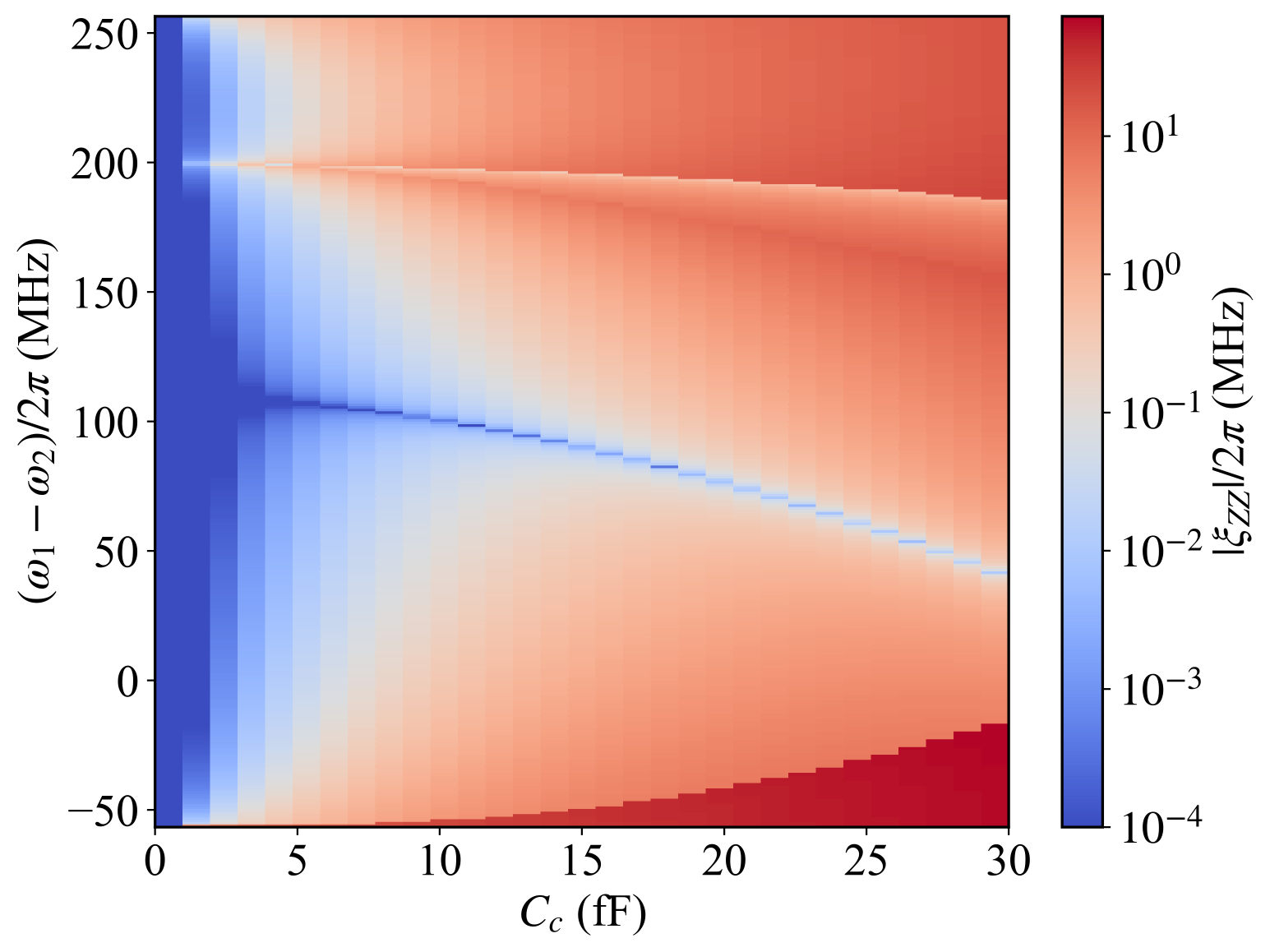}
      \put(-3,68){\normalsize (e)}
    \end{overpic}
    \label{fig:S1e}
  \end{subfigure}
  \hspace{0.1cm}
  % 子图(f)
  \begin{subfigure}[b]{0.32\textwidth}
    \centering
    \begin{overpic}[width=\linewidth]{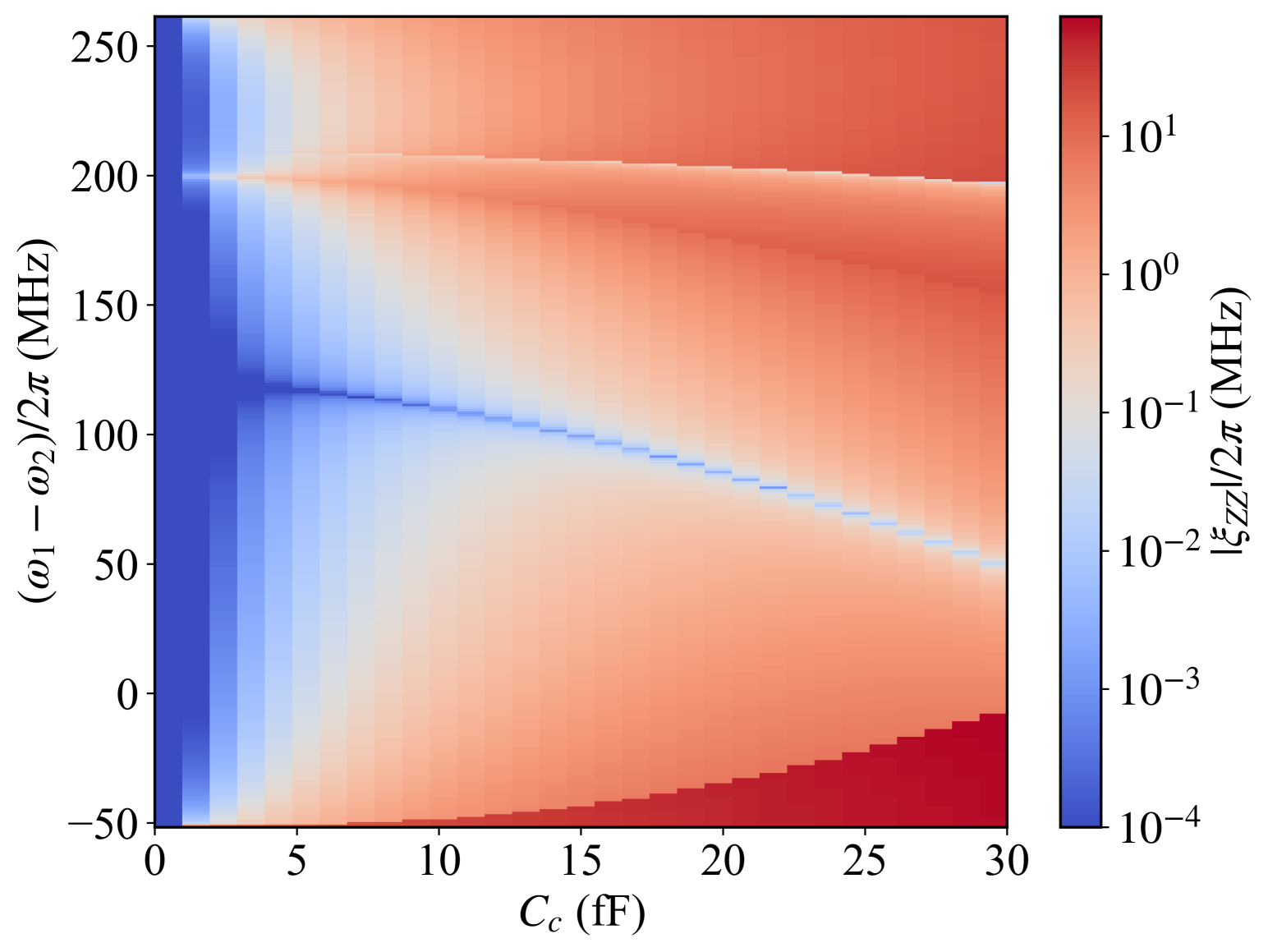}
      \put(-3,68){\normalsize (f)}
    \end{overpic}
    \label{fig:S1f}
  \end{subfigure}

  \caption{ZZ interaction strength $|\xi_{\mathrm{ZZ}}|/2\pi$ as a function of the coupling capacitance $C_{\mathrm{c}}$ and the frequency difference $(\omega_1-\omega_2)/2\pi$, with $\omega_1/2\pi$ fixed at (a) 4.620 GHz, (b) 4.625 GHz, (c) 4.630 GHz, (d) 4.635 GHz, (e) 4.640 GHz, and (f) 4.645 GHz. In each panel, a narrow region of vanishing $|\xi_{\mathrm{ZZ}}|$ exists, suitable for defining an idle point for single-qubit operations. As $\omega_1/2\pi$ increases, this idle region shifts monotonically toward larger values of $(\omega_1-\omega_2)/2\pi$. The color scale represents $|\xi_{\mathrm{ZZ}}|/2\pi$ in MHz on a logarithmic scale.}
  \label{fig:S1}
\end{figure}

To determine a suitable coupling capacitance between the qubits and the cable, we perform a two-dimensional scan of the ZZ interaction strength over the frequency difference $(\omega_1-\omega_2)/2\pi$ and the coupling capacitance $C_{\mathrm{c}}$, with $\omega_1/2\pi$ fixed at six discrete values: 4.620, 4.625, 4.630, 4.635, 4.640, and 4.645 GHz. These slices are chosen to characterize the behavior across the relevant three-dimensional parameter space while keeping the computational cost manageable. The results are shown in Fig.~\ref{fig:S1}(a)--(f). For each fixed value of $\omega_1/2\pi$, there exists a narrow region where $|\xi_{\mathrm{ZZ}}| \ll 2\pi\times10$~kHz, which is suitable for defining an idle point for single-qubit operations. As $\omega_1/2\pi$ increases, this idle region shifts monotonically toward larger $(\omega_1-\omega_2)/2\pi$, while the overall structure of the ZZ interaction remains qualitatively unchanged. Furthermore, a large ZZ interaction strength is desirable for implementing the adiabatic CZ gate, as it allows the required conditional phase to be accumulated within a shorter plateau duration. Based on these considerations, we choose $C_{\mathrm{c}}=20~\mathrm{fF}$.

\begin{figure}[htbp]
  \centering
  \begin{subfigure}[b]{0.45\textwidth}
    \includegraphics[width=\linewidth]{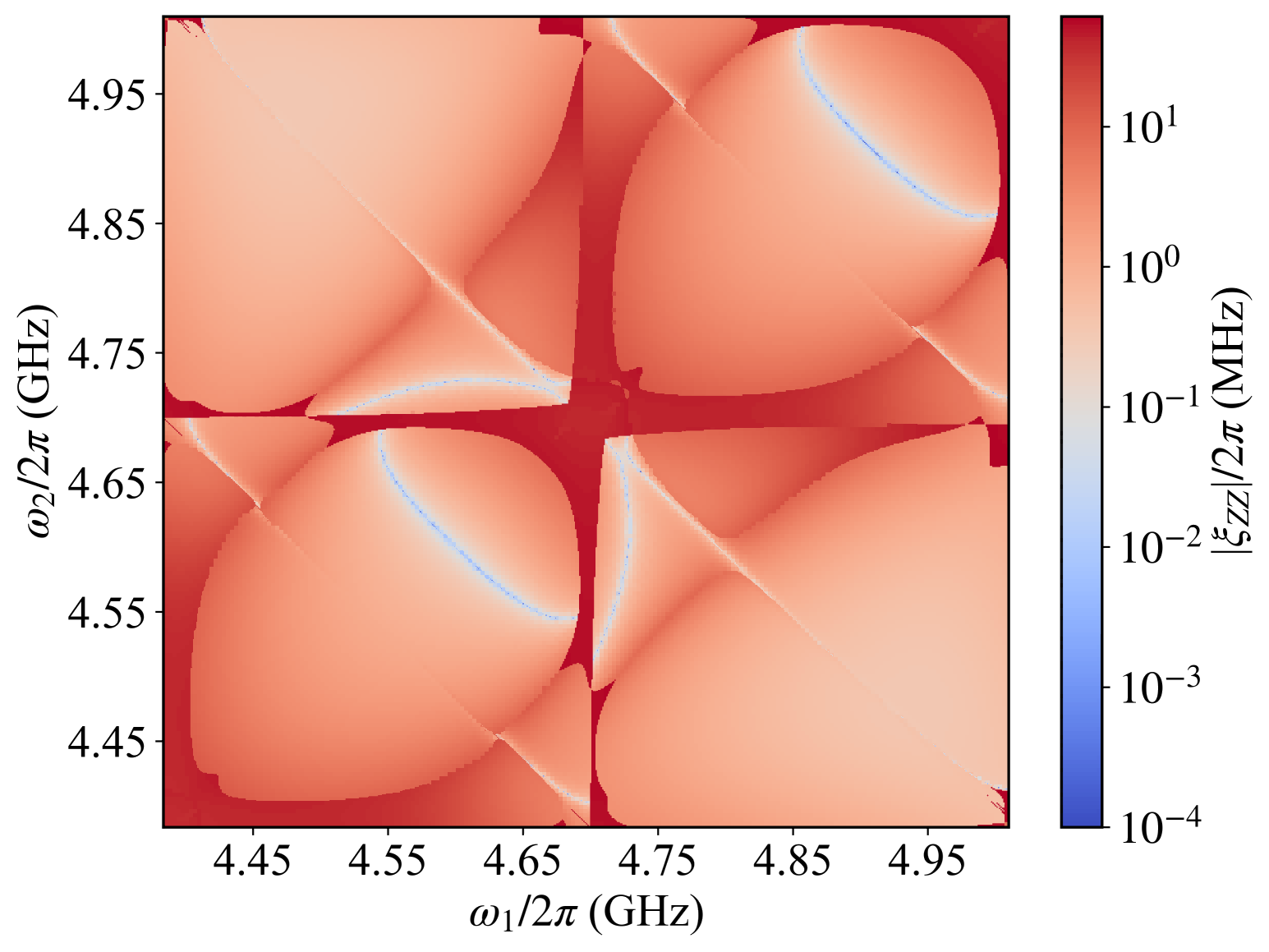}
  \end{subfigure}
  \caption{ZZ interaction for the two qubits in different mode intervals: $\omega_1/2\pi$ between modes $m=14$ and $m=15$, and $\omega_2/2\pi$ between modes $m=15$ and $m=16$. The color scale shows $|\xi_{\mathrm{ZZ}}|/2\pi$ in MHz on a logarithmic scale.}
  \label{fig:S2}
\end{figure}

Furthermore, it is also possible to place the two qubit frequencies in different mode intervals. Fig~\ref{fig:S2} presents the ZZ interaction when one qubit is tuned within the mode interval between $m=14$ and $m=15$, while the other lies between $m=15$ and $m=16$, corresponding to the upper-left and lower-right regions of the plot. This configuration breaks the diagonal symmetry of the ZZ interaction distribution. The interference between different cable modes gives rise to a rich structure of level anti-crossings and sign changes in the effective interaction. This additional flexibility provides more freedom in selecting the gate trajectory and allows access to multiple operating regimes, offering alternative routes for implementing high-fidelity remote entangling gates, in contrast to the conventional case where both qubits are tuned within the same mode interval.

\section{Working point optimization}

\begin{figure}[htbp]
  \centering
  % \captionsetup{font={footnotesize}, skip=-5pt}
  \begin{subfigure}[b]{1.0\textwidth}
    \includegraphics[width=\linewidth]{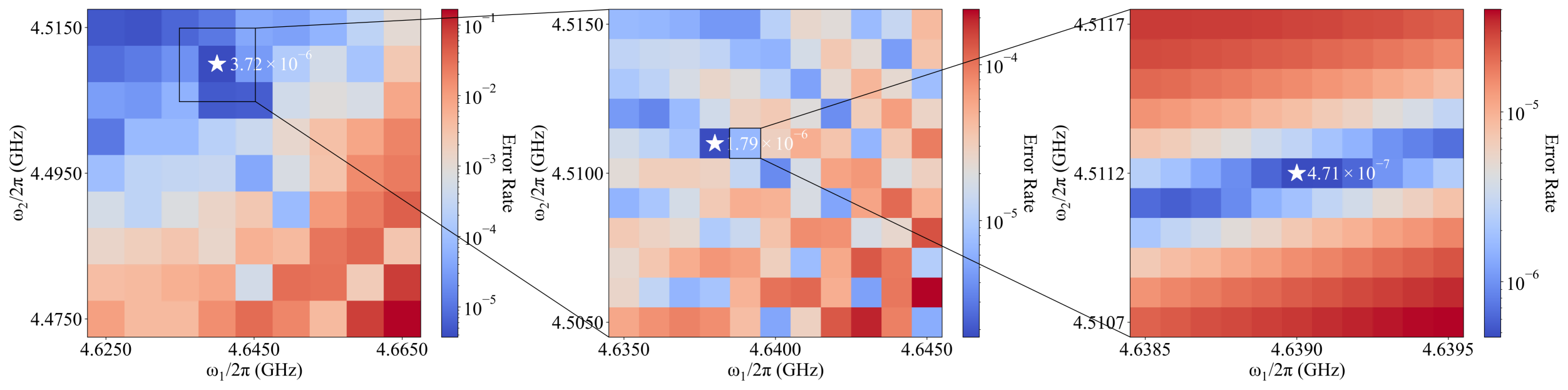}
  \end{subfigure}
  \caption{Illustration of the iterative scanning procedure for identifying the optimal working point at a cable length of 30 cm. The figure shows three successive refinement rounds: an initial coarse scan with a step size of 5 MHz over a $9\times9$ grid, followed by a 1 MHz step scan over an $11\times11$ grid, and a final fine scan with a step size of 0.1 MHz over an $11\times11$ grid. A fourth round with the same fine resolution is performed if the optimal point from the third scan lies near the boundary, ensuring that the optimum is captured. The white star marks the optimal working point ($\omega_1/2\pi = 4.6390$ GHz, $\omega_2/2\pi = 4.5112$ GHz) identified after the final refinement.}
  \label{fig:S3}
\end{figure}

To identify the optimal working point for the remote adiabatic CZ gate, we perform a two-dimensional numerical scan over the qubit frequencies $\omega_1$ and $\omega_2$. For each frequency pair, we calculate the gate fidelity as a function of the plateau duration $T_{\mathrm{plateau}}$, with the rise and fall durations fixed at 50 ns. For each working point, we construct the time-dependent Hamiltonian for the rising and falling edges, evolve the four computational basis states through the two edge segments, and evaluate the projected evolution matrix using a spectral decomposition of the plateau Hamiltonian.

The full time evolution operator for a given working point and plateau duration $t$ can be written as
\begin{equation}
U(t) = U_{\mathrm{fall}} \, U_{\mathrm{plateau}}(t) \, U_{\mathrm{rise}},
\tag{1}
\end{equation}
where $U_{\mathrm{rise}}$ and $U_{\mathrm{fall}}$ are the evolution operators for the rising and falling edges, respectively. The plateau evolution operator is decomposed spectrally as
\begin{equation}
U_{\mathrm{plateau}}(t) = e^{-iH_{\mathrm{plateau}} t} = \sum_n e^{-iE_n t} |E_n\rangle \langle E_n|,
\tag{2}
\end{equation}
where $H_{\mathrm{plateau}}$ is the time-independent Hamiltonian during the plateau, $E_n$ and $|E_n\rangle$ are its eigenenergies and eigenstates, respectively, and $n$ runs over all eigenstates retained in the numerical Hilbert space.

We then project the full evolution operator onto the dressed computational basis at the idle point, $\{|\widetilde{00}\rangle,|\widetilde{01}\rangle,|\widetilde{10}\rangle,|\widetilde{11}\rangle\}$. The resulting projected evolution matrix is
\begin{equation}
(U_{\mathrm{proj}})_{ij,kl}=\langle\widetilde{ij}|U(t)|\widetilde{kl}\rangle,
\qquad
i,j,k,l\in\{0,1\}.
\tag{3}
\end{equation}
The rows and columns of $U_{\mathrm{proj}}$ follow the basis order $|\widetilde{00}\rangle,|\widetilde{01}\rangle,|\widetilde{10}\rangle,$ and $|\widetilde{11}\rangle$.

Substituting the spectral decomposition of $U_{\mathrm{plateau}}(t)$ into Eq.~(3), we obtain
\begin{equation}
(U_{\mathrm{proj}})_{ij,kl}
=
\sum_n
\langle \widetilde{ij} | U_{\mathrm{fall}} | E_n \rangle
e^{-iE_n t}
\langle E_n | U_{\mathrm{rise}} | \widetilde{kl} \rangle.
\tag{4}
\end{equation}

Equivalently, in matrix form,
\begin{equation}
U_{\mathrm{proj}} = D \cdot \mathrm{diag}(e^{-iE_n t}) \cdot C,
\tag{5}
\end{equation}
where $D$ is a $4\times N$ matrix with elements $D_{ij,n}=\langle\widetilde{ij}|U_{\mathrm{fall}}|E_n\rangle$, and $C$ is an $N\times4$ matrix with elements $C_{n,kl}=\langle E_n|U_{\mathrm{rise}}|\widetilde{kl}\rangle$.
The diagonal matrix $\mathrm{diag}(e^{-iE_n t})$ contains the phase factors acquired by the eigenstates during the plateau.

This decomposition avoids repeated full time-propagator simulations for each plateau duration. For a fixed working point, the matrices $D$ and $C$ are independent of the plateau duration, so only the phase factors $e^{-iE_n t}$ need to be updated when scanning $T_{\mathrm{plateau}}$. After obtaining $U_{\mathrm{proj}}$, we apply virtual-$Z$ corrections to compensate the single-qubit dynamical phases. The resulting phase-corrected matrix is denoted by $U_{\mathrm{sim}}$, consistent with the definition in the main text, and is used in Eq.~(5) of the main text to evaluate the gate fidelity.

To efficiently identify high-fidelity operating points, we perform at least three to four successive rounds of scans for each cable length. This procedure provides sufficient refinement of the working point while keeping the total computation time manageable. The resulting working points are not guaranteed to be globally optimal because a complete global search over the entire frequency plane is computationally expensive, particularly in regions where the ZZ interaction is weak and a long plateau duration is required to accumulate the conditional phase. We therefore adopt three practical criteria when identifying candidate operating points: (i) the gate trajectory is chosen not to pass through the energy level anti-crossing regions; (ii) the magnitude of the ZZ interaction should exceed 1 MHz; and (iii) the exchange symmetry of the system is used to avoid redundant calculations.

\section{Performance of the gate for longer cables}

In the main text, we demonstrate the remote adiabatic CZ gate for a 30~cm coaxial cable, achieving an infidelity below \(10^{-6}\) while maintaining robustness against both qubit-frequency fluctuations and cable-length variations arising from fabrication tolerances. To further investigate the scalability of the proposed scheme to larger cryogenic systems, we extend the analysis to longer coaxial cables with lengths of 50~cm and 100~cm. The same qubit-frequency drift range of \(\pm0.5\)~MHz is considered, while the cable length is varied within a tolerance of \(\pm1\)~mm.

\begin{figure}[htbp]
  \centering
  % \captionsetup{font={footnotesize}, skip=-5pt, justification=justified, singlelinecheck=off}

  \begin{subfigure}[b]{0.35\textwidth}
    \centering
    \begin{overpic}[width=\linewidth]{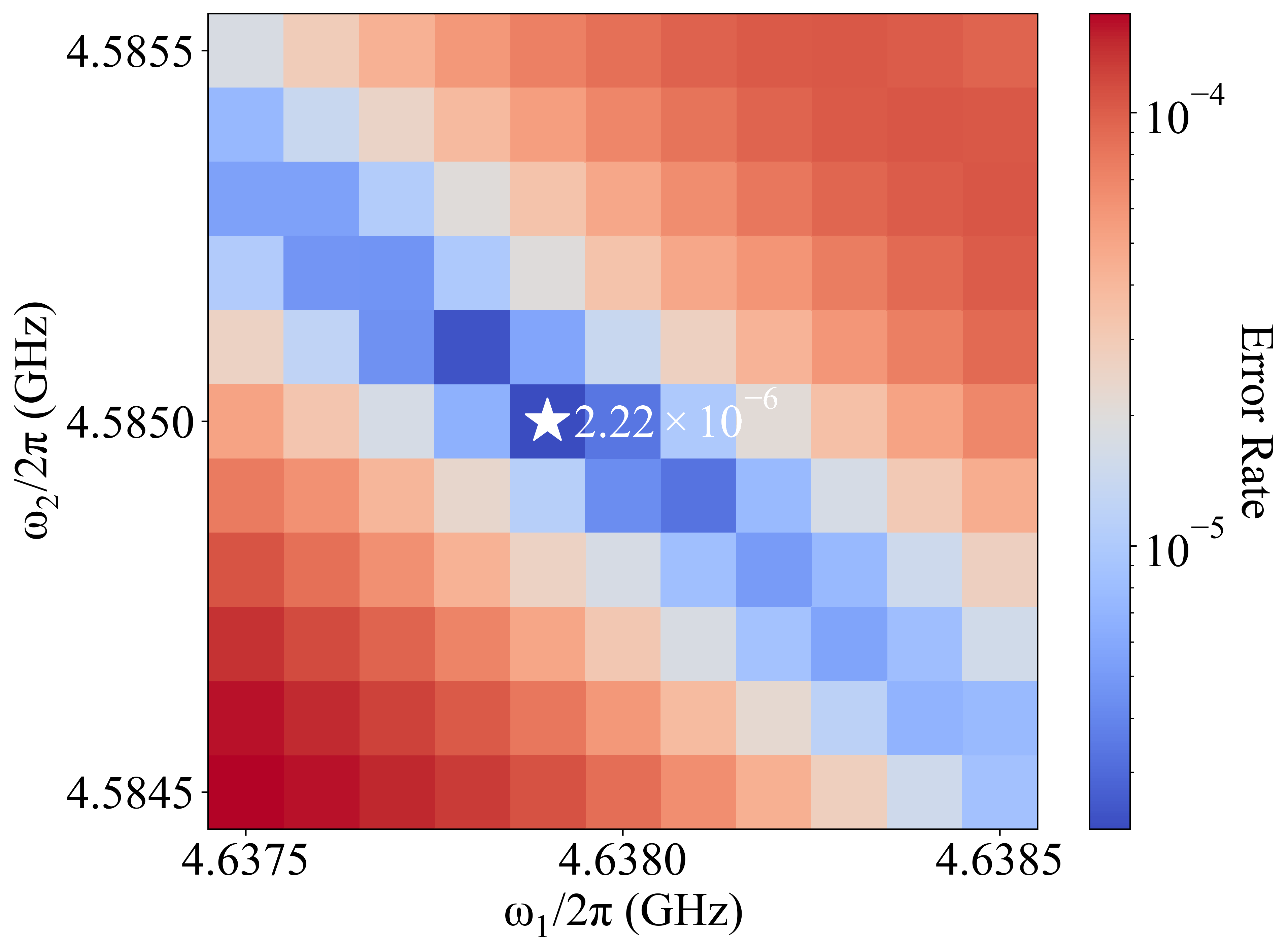}
      \put(-2.5,69){\normalsize (a)}
    \end{overpic}
    \label{fig:S4a}
  \end{subfigure}
  \hspace{0.6cm}
  \begin{subfigure}[b]{0.35\textwidth}
    \centering
    \begin{overpic}[width=\linewidth]{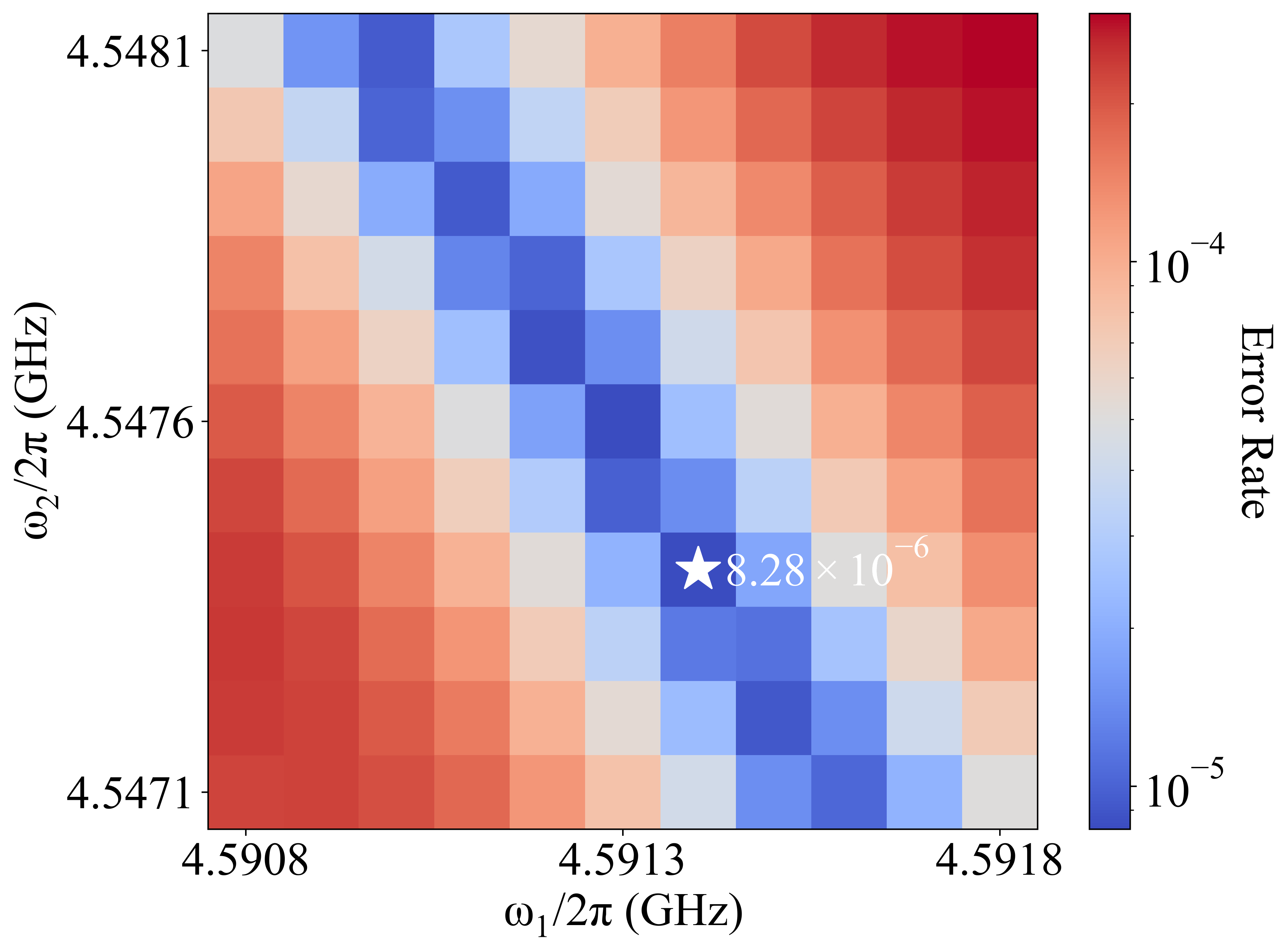}
      \put(-2.5,69){\normalsize (b)}
    \end{overpic}
    \label{fig:S4b}
  \end{subfigure}

  \begin{subfigure}[b]{0.35\textwidth}
    \centering
    \begin{overpic}[width=\linewidth]{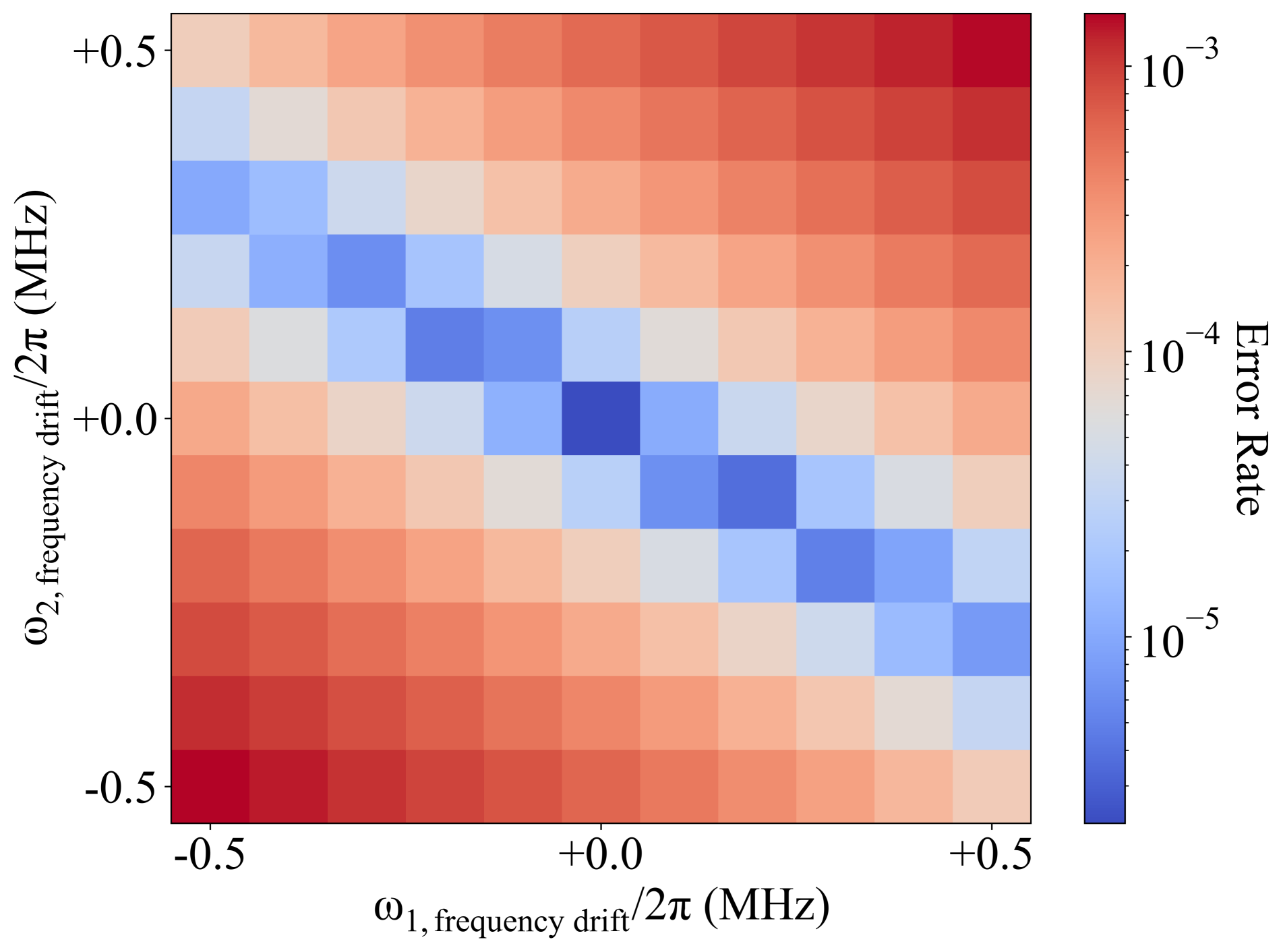}
      \put(-2.5,69){\normalsize (c)}
    \end{overpic}
    \label{fig:S4c}
  \end{subfigure}
  \hspace{0.6cm}
  \begin{subfigure}[b]{0.35\textwidth}
    \centering
    \begin{overpic}[width=\linewidth]{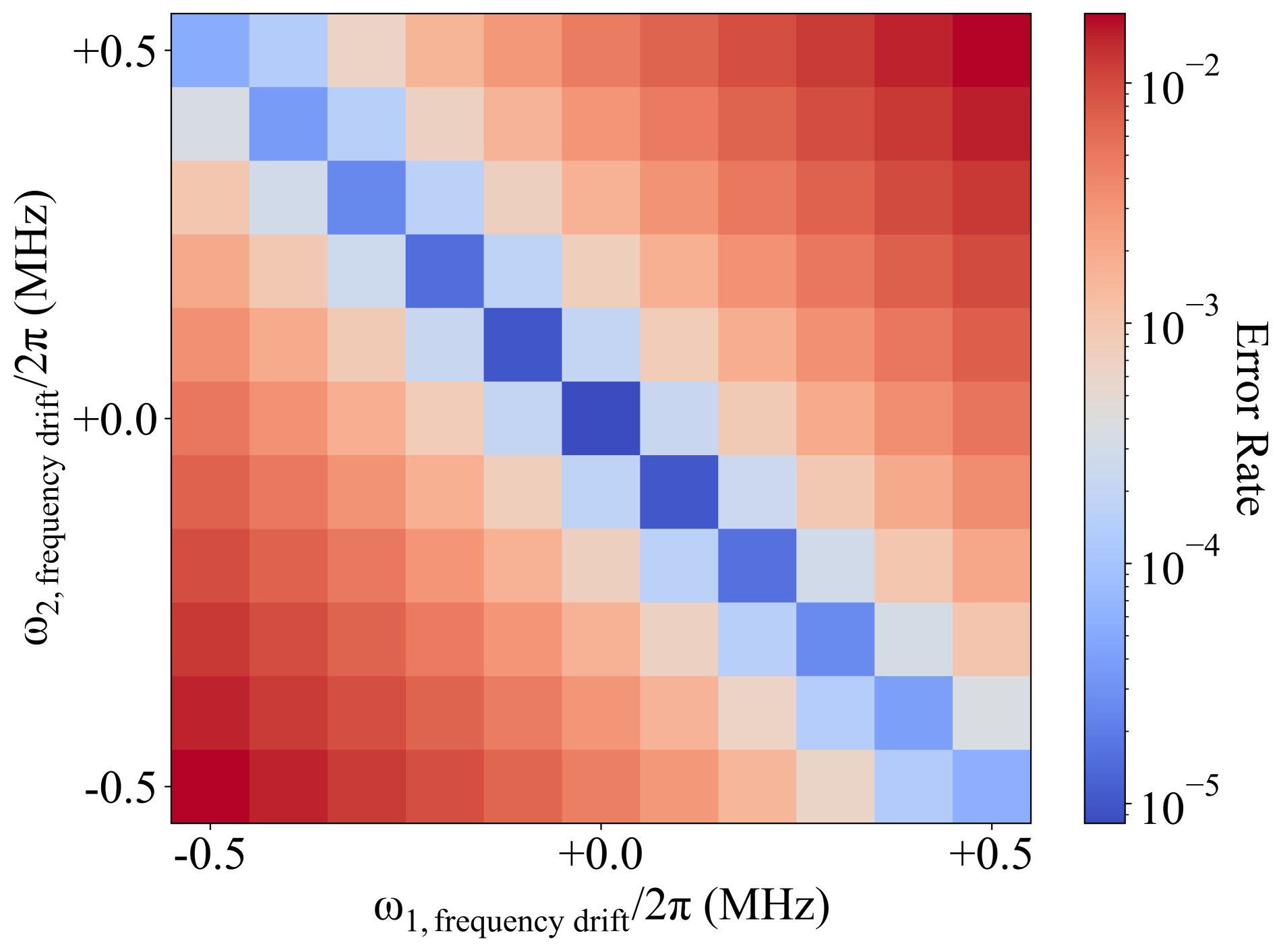}
      \put(-2.5,69){\normalsize (d)}
    \end{overpic}
    \label{fig:S4d}
  \end{subfigure}
  \caption{
  (a) Infidelity landscape used to identify the optimized working point for the remote CZ gate with a 50-cm coaxial cable over the scanned two-qubit frequency range.
  (b) Corresponding infidelity landscape for a 100-cm coaxial cable.
  (c)--(d) Gate infidelity under two-qubit frequency drift for cable lengths of 50 and 100~cm, respectively.
}
  \label{fig:S4}
\end{figure}

Figures~\ref{fig:S4}(a) and (b) show the infidelity landscapes obtained by scanning the two qubit working frequencies for cable lengths of 50 and 100~cm, respectively. These scans are used to identify suitable working points that minimize the gate infidelity for each cable length. For the 50~cm cable [Fig.~\ref{fig:S4}(a)], a minimum infidelity of approximately \(2.22\times10^{-6}\) is obtained within the scanned frequency range. For the 100~cm cable [Fig.~\ref{fig:S4}(b)], the minimum infidelity is approximately \(8.28\times10^{-6}\).

To further characterize the sensitivity of the gate to qubit-frequency drift, Fig.~\ref{fig:S4}(c)--(d) show the gate fidelity under frequency drift for cable lengths of 50 and 100~cm, respectively. The frequency drift is introduced around the optimized working frequencies, while the gate duration is fixed at the value determined at the optimized working point. At each drift point, the corresponding idle frequencies are shifted together with the working frequencies, thereby preserving the same frequency excursion of the control pulse, while the virtual-Z corrections are independently determined. Over the considered frequency-drift range, the maximum infidelities are  $1.53\times10^{-3}$, and $1.95\times10^{-2}$ for cable lengths of 50 and 100~cm, respectively.

\begin{figure}[htbp]
  \centering
  % \captionsetup{font={footnotesize}, skip=-5pt, justification=justified, singlelinecheck=off}

  \begin{subfigure}[b]{0.45\textwidth}
    \centering
    \begin{overpic}[width=\linewidth]{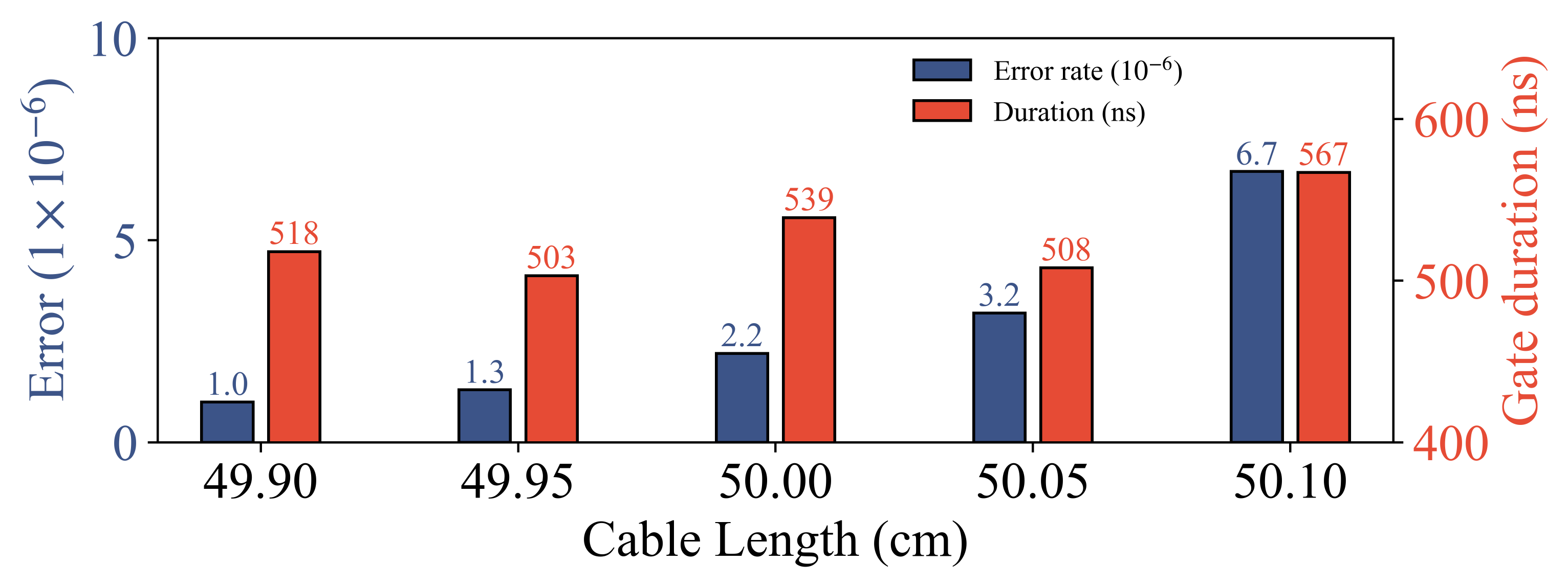}
      \put(-1.5,35){\normalsize (a)}
    \end{overpic}
    \label{fig:S5a}
  \end{subfigure}
  \hspace{0.6cm}
  \begin{subfigure}[b]{0.45\textwidth}
    \centering
    \begin{overpic}[width=\linewidth]{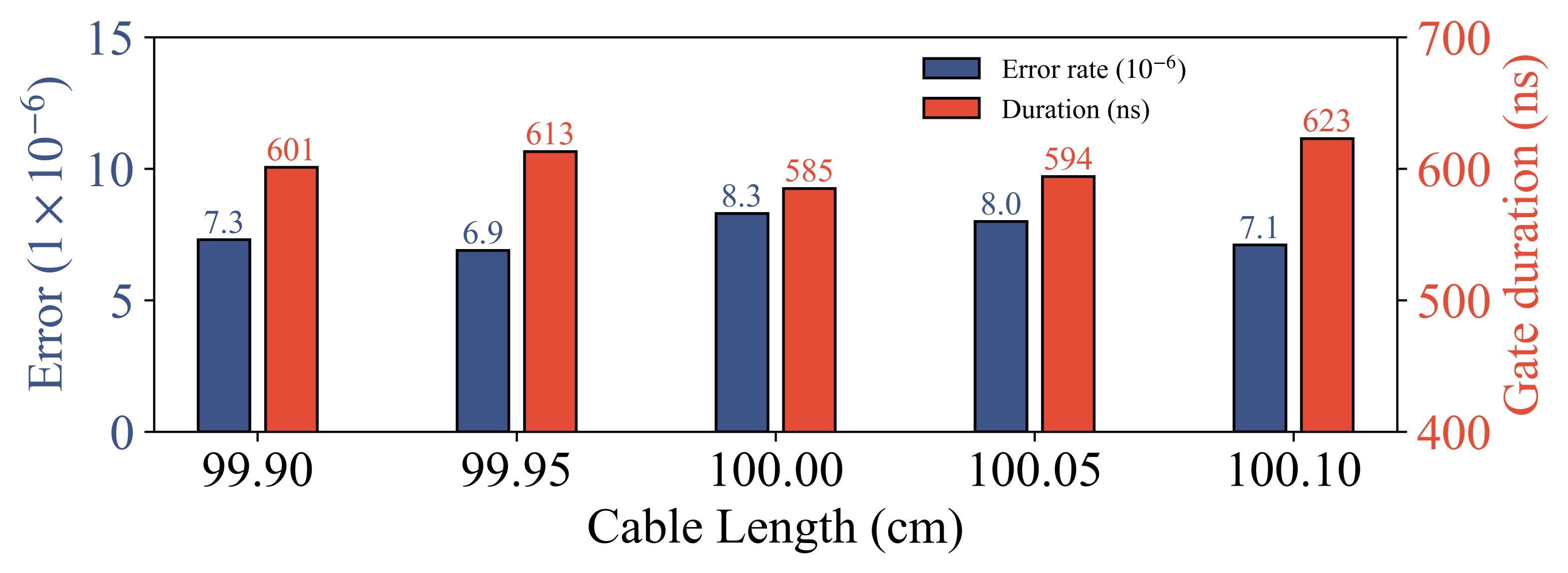}
      \put(-1.5,35){\normalsize (b)}
    \end{overpic}
    \label{fig:S5b}
  \end{subfigure}

  \caption{
  (a) Optimized infidelity and gate duration as a function of cable length for a 50-cm coaxial cable. The cable length is varied from 49.90~cm to 50.10~cm in 0.5-mm intervals, with the qubit working point re-optimized at each length.
  (b) Corresponding results for a 100-cm coaxial cable, with the cable length varied from 99.90~cm to 100.10~cm.
  }
  \label{fig:S5}
\end{figure}

We further investigate the sensitivity of the optimized gate to cable-length variations arising from fabrication tolerances. Fig.~\ref{fig:S5}(a) shows the optimized infidelity and corresponding gate duration for cable lengths ranging from 49.90~cm to 50.10~cm, sampled at 0.5-mm intervals, with the qubit working point re-optimized at each length. The optimized infidelity ranges from \(1.0\times10^{-6}\) to \(6.7\times10^{-6}\), while the gate duration ranges from 503~ns to 567~ns. For a 100~cm coaxial cable [Fig.~\ref{fig:S5}(b)], the optimized infidelity ranges from \(6.9\times10^{-6}\) to \(8.3\times10^{-6}\), with the gate duration ranging from 585~ns to 623~ns.

\section{Details of population-transfer errors}

We first summarize the definitions introduced in the main text. The instantaneous states $|\widetilde{ij}\rangle_t$ are the time-dependent eigenstates associated with the computational states $|ij\rangle$ during the gate, with $|\widetilde{ij}\rangle_0$ denoting the corresponding dressed computational state at the idle point. For a system initialized in $|\widetilde{ij}\rangle_0$, the population remaining in the corresponding instantaneous state $|\widetilde{ij}\rangle_t$ at time $t$ is denoted by $P_{ij}$, as defined in the main text, and the total population-transfer error is $\varepsilon_{ij}=1-P_{ij}$. Population transferred to a different instantaneous state $|\widetilde{kl}\rangle_t$, with $(k,l)\neq(i,j)$, is denoted by $P_{ij\rightarrow kl}$. The relevant population-transfer channels therefore satisfy $\varepsilon_{ij}=\sum_{(k,l)\neq(i,j)}P_{ij\rightarrow kl}$ within the excitation-number sector associated with the initial state.

Here, $(k,l)$ labels the two-qubit excitation states, including higher excited states such as $|2\rangle$. The cable photon occupation is determined by conservation of the total excitation number. For an initial state $|\widetilde{ij}\rangle_0$, the total excitation number is $i+j$, and thus a state with qubit excitations $(k,l)$ contains $n_{\mathrm{ph}}(k,l)=i+j-k-l$ cable photons.

We provide a mode-resolved breakdown of the population-transfer channels for the $|\widetilde{11}\rangle_0$ and $|\widetilde{10}\rangle_0$ initial states, complementing the analysis in the main text. We resolve the cable-related channels according to the specific cable modes involved, indicated by mode indices in parentheses. For example, $P_{01,(15)}$ denotes the population transferred to the state associated with $|01\rangle$ and one photon in cable mode 15, while $P_{00,(14)(15)}$ denotes the population transferred to the state associated with $|00\rangle$ with one photon in each of cable modes 14 and 15.
\begin{figure}[htbp]
  \centering
  % \captionsetup{font={footnotesize}, skip=-5pt}
  \begin{subfigure}[b]{0.464\textwidth}
    \centering
    \begin{overpic}[width=\linewidth]{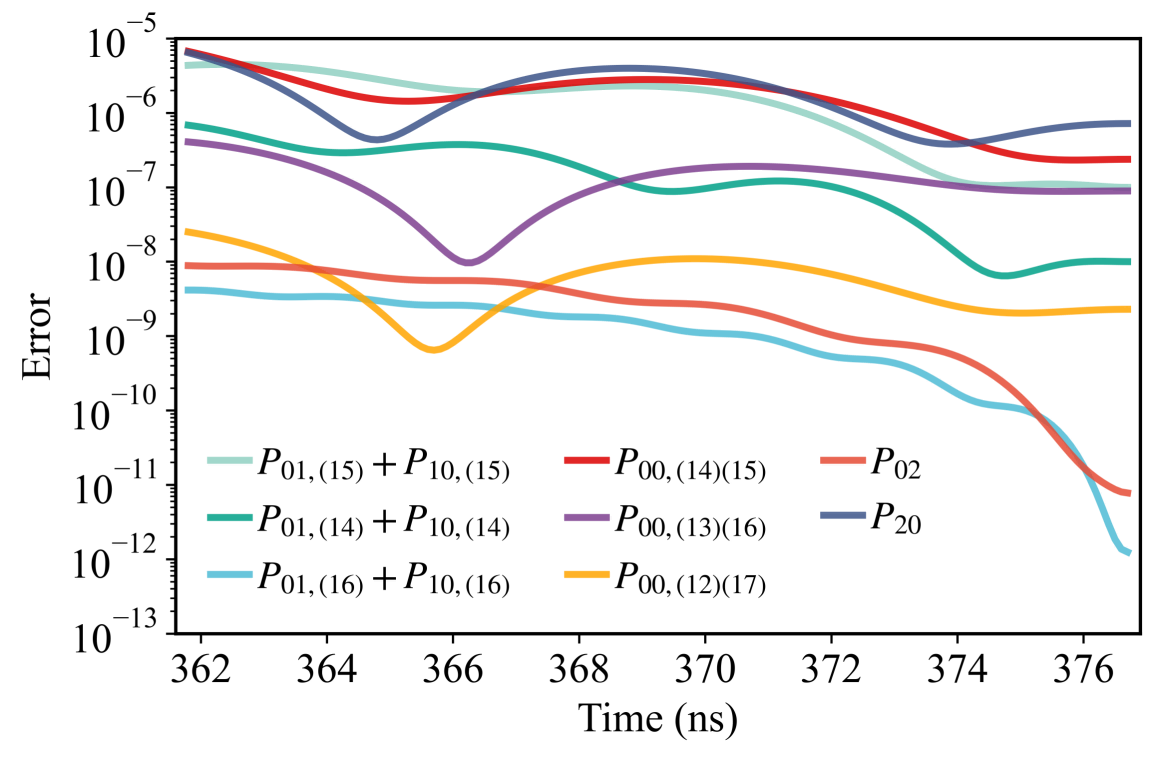}
      \put(0,60){\normalsize (a)}
    \end{overpic}
    \label{fig:S6a}
  \end{subfigure}
  \hspace{0.6cm}
  \begin{subfigure}[b]{0.45\textwidth}
    \centering
    \begin{overpic}[width=\linewidth]{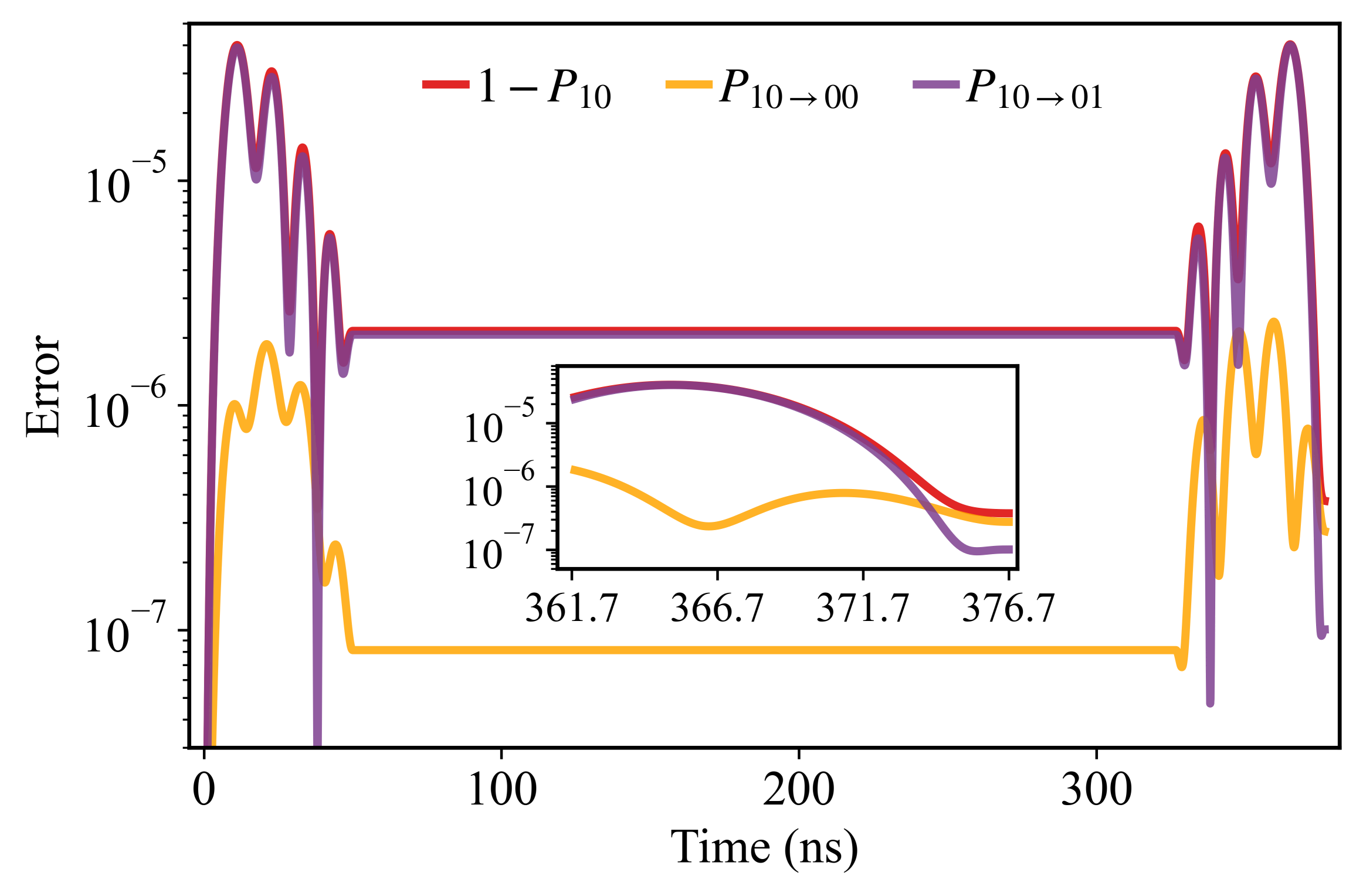}
      \put(0,60){\normalsize (b)}
    \end{overpic}
    \label{fig:S6b}
  \end{subfigure}
  \caption{(a) Detailed mode-resolved population-transfer channels during the final 15 ns for the $|\widetilde{11}\rangle_0$ initial state. The curves show the top three single-photon cable modes: $P_{01,(15)}+P_{10,(15)}$, $P_{01,(14)}+P_{10,(14)}$, and $P_{01,(16)}+P_{10,(16)}$, and the top three double-photon cable mode pairs: $P_{00,(14)(15)}$, $P_{00,(13)(16)}$, and $P_{00,(12)(17)}$. Also shown are the non-computational qubit states $P_{02}$ and $P_{20}$. The numbers in parentheses denote the cable mode indices; for example, $(15)$ denotes one photon in cable mode 15, while $(14)(15)$ denotes one photon in each of cable modes 14 and 15. (b) Population-transfer error components for the $|\widetilde{10}\rangle_0$ initial state: total error $1-P_{10}$, single-photon cable leakage, and population-swap error to the $|\widetilde{01}\rangle_t$ state. The inset shows a magnified view of the final 15 ns.}
  \label{fig:S6}
\end{figure}

Fig~\ref{fig:S6}(a) presents the mode-resolved decomposition of the population-transfer channels for the $|\widetilde{11}\rangle_0$ initial state during the final 15 ns of the gate. At the final time, the dominant residual population-transfer channel is the non-computational state $|\widetilde{20}\rangle_t$, with a population of $7.17\times10^{-7}$, while the population of $|\widetilde{02}\rangle_t$ is negligible at $7.77\times10^{-12}$. Among the single-photon cable channels, mode 15 gives the largest contribution ($9.93\times10^{-8}$), followed by mode 14 ($9.99\times10^{-9}$) and mode 16 ($1.25\times10^{-12}$). For the double-photon cable channels, the dominant contributions come from mode pairs (14,15) with $2.37\times10^{-7}$, (13,16) with $8.91\times10^{-8}$, and (12,17) with $2.29\times10^{-9}$. All other channels are at least one order of magnitude smaller. These results show that cable modes close to the qubit frequencies, particularly modes 13--16, provide the dominant contributions to photon-mediated population transfer, with the mode pair (14,15) giving the largest contribution.

For the $|\widetilde{10}\rangle_0$ initial state, shown in Fig.~\ref{fig:S6}(b), the error decomposition is similar to that of the $|\widetilde{01}\rangle_0$ initial state presented in the main text. The total population-transfer error $1-P_{10}$ reaches a final value of $3.76\times10^{-7}$, with single-photon cable leakage contributing $2.75\times10^{-7}$ and the population-swap error contributing $1.01\times10^{-7}$. Both components exhibit oscillations near the pulse edges and converge during the final 15 ns, as shown in the inset. The similarity between the error budgets of the $|\widetilde{01}\rangle_0$ and $|\widetilde{10}\rangle_0$ initial states results from the exchange symmetry of the system.

These decompositions show that the residual population-transfer error is dominated by a small number of leakage channels. For the $|\widetilde{11}\rangle_0$ initial state, the dominant contribution comes from the non-computational qubit state $|\widetilde{20}\rangle_t$, whereas single-photon cable leakage dominates the population-transfer error in the single-excitation subspace.

\section{Adiabaticity and instantaneous energy-gap analysis}

The adiabaticity of the remote CZ gate is analyzed using the instantaneous eigenstates of the time-dependent Hamiltonian. For an instantaneous eigenstate $|\mu(t)\rangle$ coupled to the $|\widetilde{11}\rangle_t$ branch, the adiabaticity factor is defined as $D_{\mu,11}(t)=|\langle \mu(t)|\dot{H}(t)|\widetilde{11}(t)\rangle|/|E_{\mu}(t)-E_{11}(t)|^2$, where $E_{\mu}(t)$ and $E_{11}(t)$ are the corresponding instantaneous eigenenergies. The total adiabaticity factor is obtained by summing the contributions from the relevant instantaneous eigenstate branches.
\begin{figure}[htbp]
  \centering

   \begin{subfigure}[b]{0.464\textwidth}
    \centering
    \begin{overpic}[width=\linewidth]{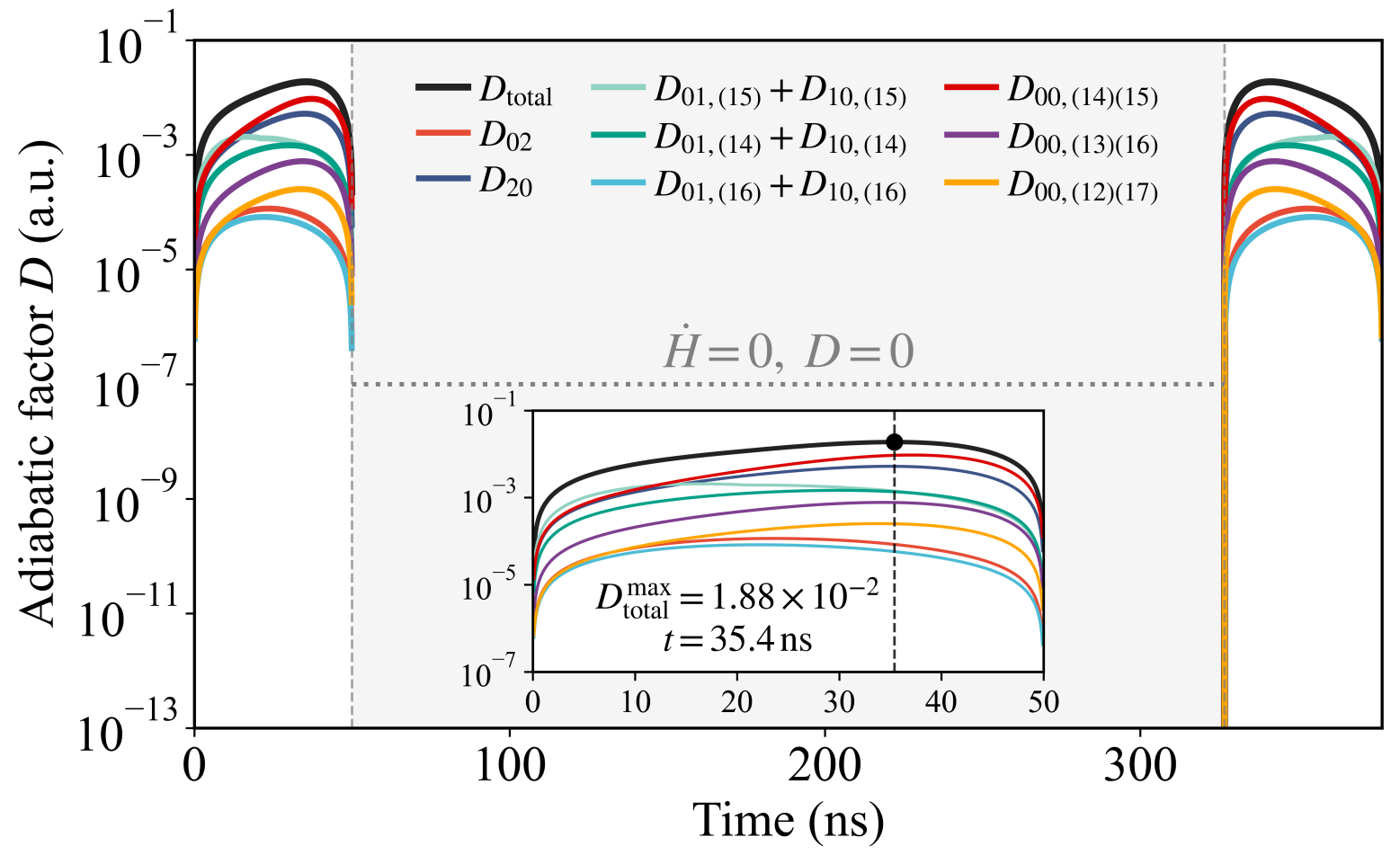}
      \put(0,56){\normalsize (a)}
    \end{overpic}
    \label{fig:S7a}
  \end{subfigure}
  \hspace{0.6cm}
  \begin{subfigure}[b]{0.45\textwidth}
    \centering
    \begin{overpic}[width=\linewidth]{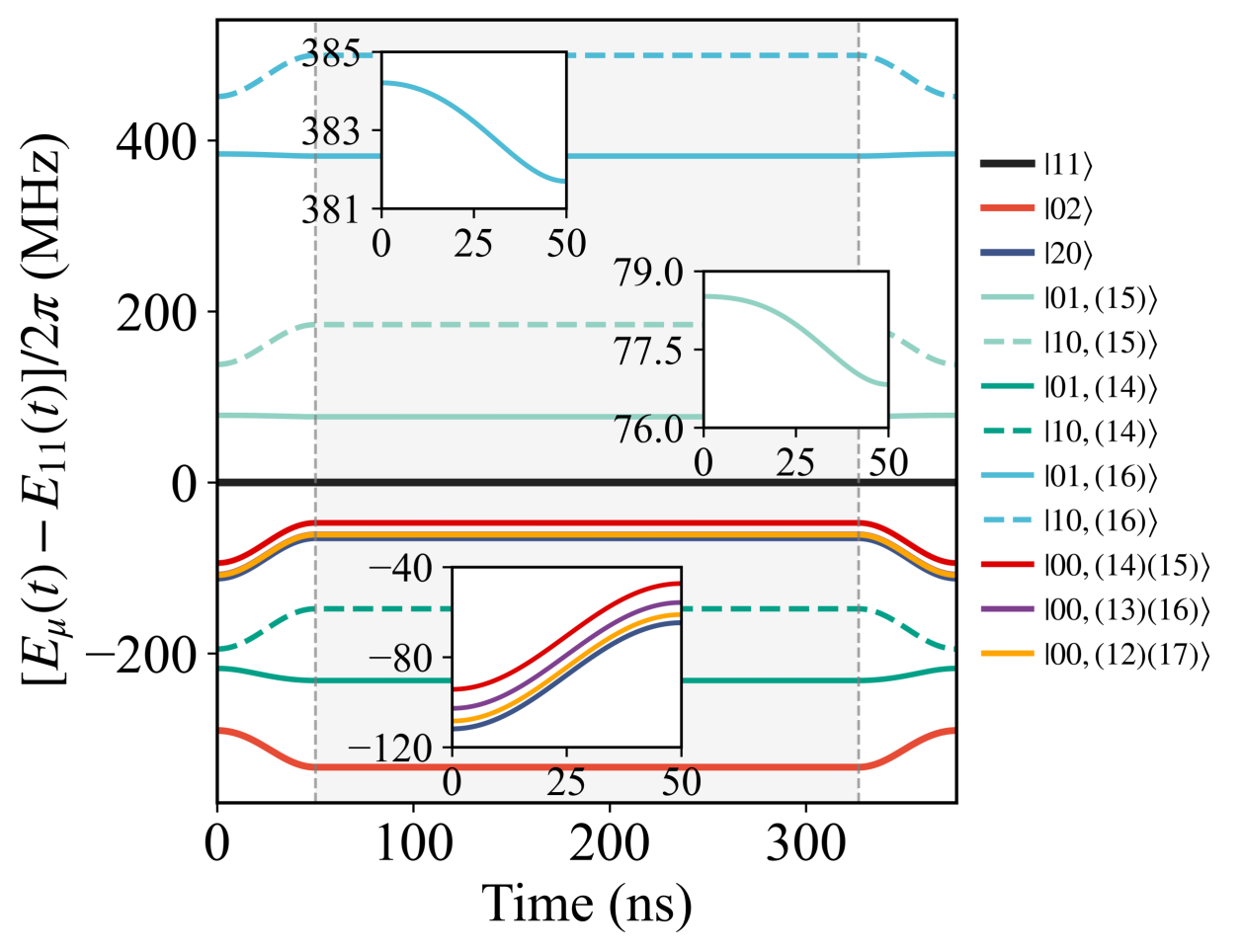}
      \put(0,72){\normalsize (b)}
    \end{overpic}
    \label{fig:S7b}
  \end{subfigure}

  \caption{
  Adiabaticity analysis of the remote CZ gate for the $|\widetilde{11}\rangle$ computational state.
  (a) Time evolution of the total adiabaticity factor $D_{\mathrm{total}}$ and the channel-resolved contributions $D_{02}$, $D_{20}$, $D_{01,(m)}+D_{10,(m)}$, and $D_{00,(m)(n)}$.
  Here, $(m)$ and $(m)(n)$ denote the cable modes involved in the corresponding transitions.
  The shaded region represents the plateau stage, where the Hamiltonian is time independent and $\dot{H}=0$.
  The inset shows the first 50 ns of the gate operation, where the maximum value of $D_{\mathrm{total}}$ is $1.88\times10^{-2}$ at $t=35.4$ ns.
  (b) Instantaneous energy-gap spectrum relative to the $|\widetilde{11}\rangle_t$ branch.
  The energy differences $[E_{\mu}(t)-E_{11}(t)]/2\pi$ are shown for the relevant states, including the qubit states $|02\rangle$ and $|20\rangle$, the single-cable-photon states $|01,(m)\rangle$ and $|10,(m)\rangle$, and the two-cable-photon states $|00,(m)(n)\rangle$.
  Here, $\mu$ labels the instantaneous eigenstate branch, while $(m)$ and $(n)$ denote the corresponding cable mode indices.
  The $|\widetilde{11}\rangle_t$ branch is taken as the reference and set to zero.
  The shaded region represents the plateau stage.
  The insets show the energy gaps during the first 50 ns.
  For visual clarity, the nearly overlapping $|00,(13)(16)\rangle$ branch is vertically offset in the corresponding inset.
  }
  \label{fig:S7}
\end{figure}

Fig~\ref{fig:S7}(a) shows the time dependence of the adiabaticity factors for the $|\widetilde{11}\rangle_t$ branch. During the rising and falling stages, the Hamiltonian varies with time and the adiabaticity factors are nonzero. During the plateau stage, the Hamiltonian is time independent, giving $\dot{H}=0$ and $D_{\mu,11}=0$. The maximum total adiabaticity factor is $D_{\mathrm{total}}^{\mathrm{max}}=1.88\times10^{-2}$ at $t=35.4$ ns, and the maximum individual channel contribution is $D_{\mathrm{max}}=9.49\times10^{-3}$.

Fig~\ref{fig:S7}(b) shows the instantaneous energy differences between the relevant eigenstate branches and the $|\widetilde{11}\rangle_t$ branch. The reference branch is set to zero, and the other branches vary with the qubit frequencies during the rising and falling stages. During the plateau stage, the Hamiltonian is time independent and the instantaneous eigenenergies remain constant. The minimum instantaneous energy gap among the plotted branches is $47.30$ MHz.

\end{document}